\documentclass[prb,letterpaper,twocolumn,floatfix,nobalancelastpage]{revtex4-1}
\usepackage{dcolumn}
\usepackage{bm}
\usepackage{amsmath}
\usepackage{bm,braket}
\usepackage{color}
\usepackage{float}
\usepackage{graphicx,subfigure}
\usepackage{amssymb}
\usepackage{sidecap}
\usepackage{float} 
\usepackage[T1]{fontenc}

\usepackage{overpic}
\usepackage{rotating}

\newcommand{\ud}{\mathrm{d}}
\newcommand{\mr}{\mathbf{r}}

\newcommand{\mG}{\mathbf{G}}

\newcommand{\mE}{\mathbf{E}}

\newcommand{\mf}{\mathbf{f}}
\newcommand{\mft}{\tilde{\mathbf{f}}}
\newcommand{\tlo}{\tilde{\omega}}

\newcommand{\me}{\mathbf{e}}

\definecolor{myBlue}{rgb}{0.0430,0.5156,0.7773}
\definecolor{light-gray}{gray}{0.95}

\newcommand{\rev}[1]{{\color{black}#1}}

\begin{document}




\title{{\em Perspective:}\\ {Modes and Mode Volumes of Leaky Optical Cavities and Plasmonic Nanoresonators}}

\author{Philip Tr{\o}st Kristensen}
\email{ptkr@fotonik.dtu.dk}
\affiliation{DTU Fotonik, Technical University of Denmark, DK-2800 Kgs. Lyngby, Denmark}
\author{Stephen Hughes}
\email{shughes@physics.queensu.ca}
\affiliation{
Department of Physics, Engineering Physics and Astronomy, Queens University, Kingston, Ontario, Canada K7L 3N6}

\begin{abstract}
Electromagnetic cavity modes in photonic and plasmonic resonators offer rich and attractive regimes for tailoring the properties of light-matter interactions. Yet there is a disturbing lack of a precise definition for what constitutes a cavity mode, and as a result their mathematical properties remain largely unspecified. The lack of a definition is evidenced in part by the diverse nomenclature at use---``resonance,'' ``leaky mode,''  ``quasimode,'' to name but a few---suggesting that the dissipative nature of cavity modes somehow makes them different from other modes, but an explicit distinction is rarely made. This perspective article aims to introduce the reader to some of the subtleties and working definitions that can be rigorously applied when describing the modal properties of leaky optical cavities and plasmonic nanoresonators. We describe some recent development in the field, including {calculation methods for quasinormal modes of both photonic and plasmonic resonators and the concept of a generalized effective mode volume, and we} illustrate the theory with several representative cavity structures from the fields of photonic crystals and nanoplasmonics.

\end{abstract}

\maketitle

\section*{{Introduction}}
Optical cavities \cite{ChangCampillo1996}, and their associated cavity \emph{modes}, are ubiquitous in both classical and quantum optics and they are largely responsible for the development of semiconductor cavity \rev{quantum electrodynamics (QED)}  \cite{Reithmaier_Nature_42_197_2004,Yoshie_Nature_432_200_2004, Press_PRL_98_117402_2007} and microcavity  lasers \cite{Vahala_Nature_424_839_2003}. With a continuing drive towards miniaturization and nanophotonics, researchers are now exploring nanoscale cavity systems in more complex geometries, including {plasmonic} nanoresonators~\cite{Bergman_PRL_90_027402_2003, Odom_NL_12_5769_2012, Maier_2007, Noginov_Nature_460_1110_2009, Novotny_NaturePhot_5_83_2011, Chang_NaturePhys_3_807_2007, Kusar_APL_100_231102_2012, Andersen_NaturePhys_7_215_2011, Tame_NaturePhys_9_329_2013, BeriniNatureLaser}.
Plasmonic systems offer an attractive alternative to dielectric cavity systems since the optical fields can be confined in much smaller geometries \cite{Bergman_PRL_90_027402_2003, Odom_NL_12_5769_2012, Maier_2007, Noginov_Nature_460_1110_2009, Novotny_NaturePhot_5_83_2011, Chang_NaturePhys_3_807_2007, Kusar_APL_100_231102_2012, Andersen_NaturePhys_7_215_2011, Tame_NaturePhys_9_329_2013, BeriniNatureLaser}. 
For both dielectric cavity systems and plasmonic nanoresonators, rapid progress has been made over the last decade. For example, strong coupling with single quantum dots has been  observed in various semiconductor cavity systems \cite{Reithmaier_Nature_42_197_2004,Yoshie_Nature_432_200_2004, Press_PRL_98_117402_2007}, and \rev{Belacel} {\em et al.} \cite{Belacel_NL_13_1516_2013} have experimentally demonstrated control of the spontaneous emission rate of colloidal quantum dots (QDs) deterministically positioned in a plasmonic patch antennas. \rev{As a relatively new application, cavity optomechanics is a} branch of cavity physics that has been developing at a tremendous rate\cite{AnetsbergerOptoMech, ThompsonOptMech, PainterOptoMech1, KippenbergOptoMech1}. \rev{Theoretically, rich} quantum optical regimes, such as the asymmetric Mollow triplet, have been predicted for coherently excited QD plasmonic systems \cite{GePRB2013}, and for small separation distances between the emitter and a metal particle, the strong coupling regime has been predicted for QDs at room temperature~\cite{HohenesterPRB2008, SavastaNano, ColePRBStrongCoupling}.


\begin{figure}[t!]
\includegraphics[width=\columnwidth]{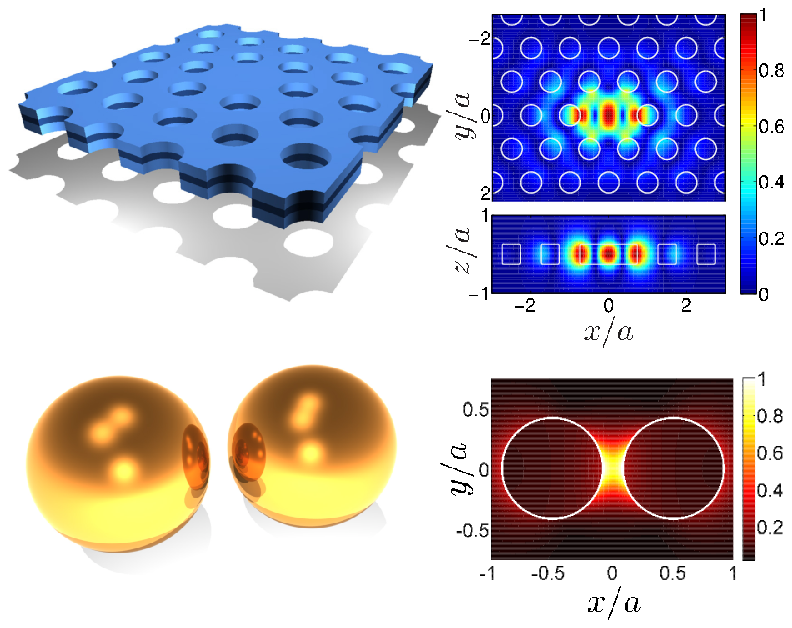}
\caption{\label{Fig:3DsketchPlusModes}Two different resonant electromagnetic material systems with examples of resonant modes. Top (left):   Photonic crystal constructed from a triangular lattice of air holes (lattice constant $a$) in a membrane of high refractive index. A defect cavity is formed by the omission of a single hole; top (right): absolute value of the cavity mode in the planes $z=0$ (top) and $y=0$ (bottom). {Figure from Ref.~\onlinecite{Kristensen_OL_37_1649_2012}.} Bottom (left): Plasmonic dimer made from two metal spheres (center to center distance $a$) in a low-index background; bottom (right): absolute value of the bright dipole mode in the $z=0$ plane through the center of the spheres. \rev{This mode was calculated as in Ref.~\onlinecite{deLasson_JOSAB_30_1996_2013} using material parameters from the text.} 
}
\end{figure}

In spite of the widespread use and exploitation of cavity modes, there appears to be 
no common consensus of a rigorous definition. Consequently, 
many of their mathematical properties, as currently in use throughout the literature, are ambiguous {or} ill-defined. The lack of a precise definition seems to be a common problem to both \rev{optical} cavities and plasmonic nanoresonators as
well as hybrid systems made from dielectric cavities with metal particles \cite{Barth, Gordon12}. Figure \ref{Fig:3DsketchPlusModes} shows two commonly studied \rev{resonant} structures: ($i$) a planar photonic crystal cavity made from a dielectric 
membrane, and ($ii$) a plasmonic dimer structure made from two spherical metal particles. Both of these cavity systems can significantly enhance light-matter interactions by trapping light at the cavity mode frequencies, and the physics of such light-matter interactions can, with care, be conveniently described in terms of the resonant cavity modes. But what exactly is a cavity mode? We argue that most, if not all, confusion about the cavity modes of general photonic and plasmonic resonators can be removed by a proper treatment within the framework of quasinormal modes (QNMs)~\cite{Lai_PRA_41_5187_1990, Leung_PRA_49_3057_1994, Ching_1996, Leung_JOSAB_13_805_1996, Lee_JOSAB_16_1409_1999, Lee_JOSAB_16_1418_1999}. Quasinormal modes are fundamentally different from the modes of most introductory textbooks on optics; they appear as solutions to a non-Hermitian differential equation problem with complex eigenfrequencies and, consequently, many familiar concepts derived for the normal modes of Hermitian problems do not apply \cite{Lee_JOSAB_16_1409_1999}. The Mie resonances~\cite{Mie, Maier_2007} in dielectric microdroplets, which are also known as morphology-dependent resonances~\cite{Lai_PRA_41_5187_1990}, are well known examples of QNMs. In general, however, all resonances with a finite quality value $Q$ can be associated with a QNM. \rev{For dielectric optical cavities, the $Q$ values quantify the leaky nature of the cavity mode. In metallic resonators, the fields not only leak out; in this case the Q values are further reduced due to absorption.} 
Although not widely appreciated in the broader nanophotonics community, QNMs have been used \rev{in modeling of complex or random lasers~\cite{Tureci_PRA_74_043822_2006, Andreasen_AdvOptPhot_3_88_2011} as well as for investigations of transmission\cite{Settimi_PRE_68_026614_2003, Settimi_JOSAB_26_876_2009} and coupled cavities\cite{Maksimovic_OptEng_47_114601_2008} in one-dimensional photonic crystals. Moreover, QNMs have been employed {by some authors} as starting points for quantized theories of optical cavities~\cite{Ho_PRE_58_2965_1998, Dutra_PRA_62_063805_2000, Severini_PRE_70_056614_2004, Dignam_PRA_85_013809_2012} and for studying quantum properties of dipole emitters in coupled cavity systems~\cite{MarcQM1,MarcQM2}.} 
Similar \rev{modes} 
are known in electronic scattering problems where the \rev{electron} states leak out, yielding {so-called} ``Siegert states''~\cite{Siegert39,Siegert05}. As \rev{optical} and plasmonic cavity structures become more complicated, it is of increasing importance to have a solid grasp of the 
associated resonant modes. 

In this perspectives article, we do not attempt to give a review of optical cavities which can be found in many excellent articles elsewhere. Rather, we describe some recent developments in the numerical calculation of QNMs and the application of these modes as a rigorous mathematical framework for understanding \rev{the electromagnetic response} of resonant systems. \rev{We first provide a rigorous definition of QNMs and discuss various calculation methods as well as the non-trivial inner product used for normalization. In addition, we discuss how the QNMs differ from the normal modes of typical introductory textbooks and remark on the use of scattered fields as approximations to QNMs. Next, we elaborate on the need to introduce a {\em generalized} effective mode volume and its use in Purcell factor calculations for optical cavities, and we highlight the difficulties associated with an extension of the formalism to plasmonic material systems. Last, we discuss how a QNM approach relates to alternative modeling schemes and list a number of possible future applications of QNMs in nanophotonics modeling.}


\section*{{Definition and practical calculation of quasinormal modes}}
\begin{figure}[htb]
\includegraphics[width=\columnwidth]{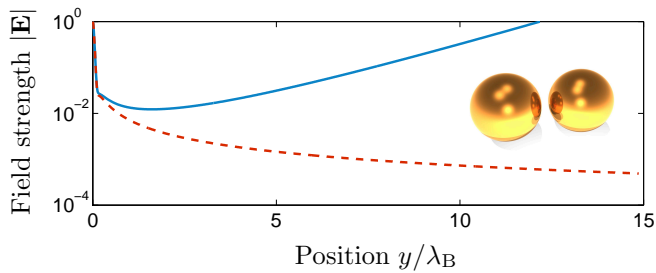}
\caption{\label{Fig:plasmonicDimerModeDivergence}{Absolute value of the fields along the line $x=z=0$ \rev{perpendicular to the dimer axis} for the plasmonic dimer mode in Fig.~\ref{Fig:3DsketchPlusModes} (solid) and  the scattered field when illuminated by a plane wave \rev{incident} along the $y$-axis \rev{and polarized along the dimer axis} (dashed).} 
}
\end{figure}

We define the electromagnetic modes of localized resonators, be they photonic, plasmonic or hybrid, as time-harmonic solutions to the source-free Maxwell equations of the form
\begin{align}
\mE(\mr,t) = \mE(\mr,\omega)\exp\{-\text{i}\omega t\},
\label{Eq:timeHarmonicSolution}
\end{align}
where the position dependent field ${\bf E}(\mr,\omega)$ solves the wave equation
\begin{align}
\nabla\times\nabla\times \mE(\mr,\omega)-k^2\epsilon_\text{}(\mr,\omega)\mE(\mr,\omega) = 0,
\label{Eq:HelmholtzEq}
\end{align}
in which $\epsilon_\text{}(\mr, \omega)$ is the position and frequency dependent relative permittivity 
and $k=\omega/c$ is the ratio of the angular frequency to the speed of light in vacuum. The wave equation alone, however, can never provide any meaningful definition of a mode---only by specifying a suitable set of boundary conditions do we get a differential equation problem with corresponding solutions that we might define as the modes. The choice of boundary condition should reflect the kind of physics one is trying to model. 
For localized resonators embedded in an otherwise homogeneous permittivity distribution $\epsilon_\text{B}=n_\text{B}^2$, {the proper} choice of boundary condition is the Silver-M\"uller radiation condition~\cite{Martin_MultipleScattering},
\begin{align}
\rev{\hat{\mr}\times\nabla\times\mE(\mr,\omega) + \text{i}n_\text{B}k\mE(\mr,\omega) \rightarrow 0\quad\text{as}\;|\mr|\rightarrow\infty,}
\label{Eq:SilverMuller}
\end{align}
where $\hat{\mr}$ is a unit vector in the direction of $\mr$. \rev{We note that Eq.~(\ref{Eq:SilverMuller}) is also known as the Sommerfeld radiation condition, in particular for scalar fields.} The use of a radiation condition turns Eq.~(\ref{Eq:HelmholtzEq}) into a non-Hermitian eigenvalue problem, even if ${\epsilon_\text{}(\mr,\omega)}$ is real. The eigenmodes are QNMs 
$\mft_\mu(\mr)$ with a discrete spectrum of complex resonance frequencies $\tlo_\mu=\omega_\mu-\text{i}\gamma_\mu$\rev{, where $\gamma_\mu>0$,} from which the $Q$ value can be calculated as $Q=\omega_\mu/2\gamma_\mu$. The radiation condition ensures that light propagates away from the cavity as expected for a leaky resonator, but this comes at the price of a conceptually challenging property of the QNMs, namely the fact that they diverge (exponentially) \rev{at large distances.} 
Although not widely appreciated, this divergence is a direct consequence of the radiation condition in Eq.~(\ref{Eq:SilverMuller}) \rev{in connection} with a complex resonance frequency. In Fig.~\ref{Fig:plasmonicDimerModeDivergence} we show this divergence explicitly for the metallic dimer in Fig.~\ref{Fig:3DsketchPlusModes}.



The QNMs can be calculated analytically for sufficiently simple structures, 
but in general one must use numerical methods. Although a great deal of electromagnetic mode solvers are available, most of them are not immediately compatible with the radiation condition which is defined only in the limit $|\mr|\rightarrow\infty$. For this reason, the use of perfectly matched layers (PMLs) is often the method of choice for practical calculations. Using PMLs, or by applying Eq.~(\ref{Eq:SilverMuller}) at the edge of the (finite) calculation domain, the QNMs may be calculated as the solutions to Eq.~(\ref{Eq:HelmholtzEq}) with typical frequency domain methods such as finite differences or finite elements~\cite{Iserles_2003}. Another frequency domain method is the \rev{(aperiodic) 
Fourier Modal Method~\cite{Armaroli_JOSAA_25_667_2008}}, also known as rigorous coupled wave analysis~\cite{Moharam_JOSA_71_811_1981}, in which the geometry is divided into subsections with known solutions. These solutions are subsequently combined to the full solution using a scattering matrix formalism, and the QNM frequencies appear as the poles of the scattering matrix. An alternative option which avoids the use of PMLs, is to calculate the QNMs from a Fredholm type integral equation~\cite{Kristensen_OL_37_1649_2012, deLasson_JOSAB_30_1996_2013} in which case the radiation condition is perfectly fulfilled by construction. \rev{This approach was used to calculate the plasmon mode in Fig.~\ref{Fig:3DsketchPlusModes}}. 
A popular alternative to the frequency domain methods is based on the fact that if the mode of interest leaks relatively slowly from the cavity, then it will be the dominant field in the cavity at long times after an initial short excitation. This means that one can calculate the QNMs using time-domain approaches, such as the well established finite-difference time-domain (FDTD) method~\cite{Tavlove_1995} in connection with a run-time Fourier transform for obtaining the spatial \rev{variation} of a QNM at select frequencies. This approach, however, may suffer from difficulties in exciting only a single QNM,
{which may be of particular concern for} 
plasmonic systems where the $Q$ values are typically relatively low. In these cases one should thoroughly analyze the scattering spectra from different excitations to confirm that only a single mode is in the bandwidth of interest. Kristensen {\em et al.}~\cite{Kristensen_OL_37_1649_2012} explicitly shows that the mode profiles as calculated using FDTD with PMLs agree with the Fredholm integral equation approach, even at large distances, where the exponential divergence sets in. The fact that the eigenvalue enters in the boundary condition makes the solution of Eq.~(\ref{Eq:HelmholtzEq}) a non-linear problem, and the precise calculation of the complex resonance frequencies is a difficult numerical task in general. Maes {\em et al}.~\cite{Maes_OE_21_6794_2013} compares solutions of a coupled cavity-waveguide system using four different electric field solvers and shows a rather large variation; in particular for the calculated $Q$ values. Last, we note that in addition to full numerical solutions, a number of powerful approximate approaches based on generalized Fabry-P\'erot models have been used for both photonic crystal cavities~\cite{Lalanne_LaserPhotRev_2_514_2008} and plasmonic nanorods~\cite{Bryant_NL_8_631_2008, Taminiau_NL_11_1020_2011}.

Because of the divergent behavior of the fields, the QNMs are non-trivial to normalize. In dispersive materials, the proper generalization of the inner product is~\cite{Leung_PRA_49_3982} 
\begin{align}
\langle\langle\mft_\text{c}|\mft_\text{c}\rangle\rangle = &\lim_{V\rightarrow\infty}\int_V \sigma(\mr,\tlo)\mft_\text{c}(\mr)\cdot\mft_\text{c}(\mr)\ud\mr \nonumber \\&+\text{i}\frac{n_\text{B}c}{2\tlo_\text{c}} \int_S \mft_\text{c}(\mr)\cdot\mft_\text{c}(\mr)\ud\mr,
\label{Eq:innerProdQuasiNormalModes}
\end{align}
where
\begin{align}
\sigma(\mr,\omega)=\frac{1}{2\omega}\frac{\partial(\epsilon_\text{}(\mr,\omega)\,\omega^2)}{\partial\omega}.
\end{align}
Both terms in Eq.~(\ref{Eq:innerProdQuasiNormalModes}) diverge, but the sum remains finite. Leung {\em et al.}~\cite{Leung_PRA_49_3982} introduced  this useful normalization to a lossy one-dimensional cavity in 1994, which we have extended above to three-dimensional problems. Recently it was pointed out by Sauvan \emph{et. al.} \cite{Sauvan_PRL_110_237401_2013} that the use of coordinate transforms (with PMLs) can dramatically improve the evaluation of the inner product when formulated as a single integral. \rev{The question of completeness of QNMs has been proven explicitly for positions within the outermost surfaces of discontinuity of the permittivity distribution in one-dimensional systems as well as in spherically symmetric material systems~\cite{Leung_JOSAB_13_805_1996, Lee_JOSAB_16_1409_1999}. To the best of our knowledge, however, there is no proof of completeness for general permittivity distributions. Nevertheless, direct application of the formalism to non-spherical material systems result in impressively good approximations, so it seems reasonable to assume completeness also for more general geometries of practical interest. Although the question of completeness is of 
formal importance, in many practical applications 
one can always approximate the electric field using only a single or a few QNMs. The neglect of all other QNMs then by construction results in formally uncontrolled approximations, but has the important quality that the resulting expressions become physically transparent and directly amenable to analytical treatment.}


As discussed above, the QNMs appear as solutions to the wave equation when imposing the Silver-M\"uller radiation condition. It is instructive to compare this choice of boundary condition to the typical choice in textbooks. In most introductory discussions about modes, it is customary to  consider localized or (discrete) translationally invariant material systems for which Dirichlet or periodic boundary conditions are appropriate. This is the case, for example for many analyses of optical waveguides. {Assuming a lossless dielectric structure}, in this case the eigenvalue problem is Hermitian and the solutions are normal modes with real eigenfrequencies that we write as $\mathbf{f}_\mu(\mr)$ and $\omega_\mu$, respectively. The normal modes are typically normalized by the inner product
\begin{align}
\langle\mf_\mu|\mf_\lambda\rangle = \int_V\epsilon_\text{}(\mr)\, \mf^*_\mu(\mr)\cdot\mf_\lambda(\mr)\,\ud\mr, 
\label{Eq:normalizationOfNormalModes}
\end{align}
where the integral is over the volume defined by the boundaries. In many applications the limit $V\rightarrow\infty$ is taken in which case the spectrum of eigenvalues becomes continuous.
\rev{
Although normal modes are often used to analyze optical waveguides, it is well known that they give rise also to so-called leaky modes\cite{SnyderLoveBook}, which diverge at large distances in the same way as QNMs. The divergence introduces normalization problems in much the same way as for QNMs, and therefore the use of leaky modes is sometimes avoided by phrasing the entire problem in terms of coupling to normal modes of the environment. For metallic waveguides, for example, Breukelaar {\em et al.}\cite{Berini2006} have used a normal mode  method  to model the radiative spreading of surface plasmon-polariton  modes into regions where the bound surface mode is cut off or radiative, and found good agreement with experiments. A similar approach is not possible for resonant cavity systems, since the QNMs are inherently leaky and there is no obvious way of defining the coupling between QNMs and the normal modes of the environment. 
}


%

\begin{figure}[htb]
\includegraphics[width=\columnwidth]{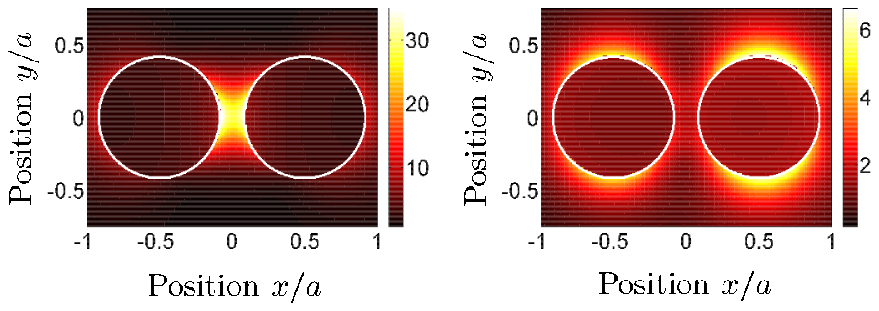}
%
\caption{\label{Fig:plasmonicDimerScatteredFields}Absolute value of the scattered field close to the dimer in Fig.~\ref{Fig:3DsketchPlusModes} when illuminated at the dipole mode resonance frequency by plane waves \rev{of unit amplitude} along the $y$-axis (left) and $x$-axis (right).}
\end{figure}

{For completeness, we remark also on the practice of calculating resonant modes from scattering calculation by subtracting the incident field. Clearly, most scattering calculations respect the radiation condition, but the scattered field depends sensitively on the choice of excitation as we show explicitly in Fig.~\ref{Fig:plasmonicDimerScatteredFields} for the case of the plasmonic dimer of Fig.~\ref{Fig:3DsketchPlusModes}. In addition, the use of an incident field means that the system is driven at a real frequency, so the resulting scattered field fails to show the expected divergence at large distances, as illustrated in Fig.~\ref{Fig:plasmonicDimerModeDivergence}, and there is no (known) meaningful way of normalizing it.} \rev{Nevertheless, comparing Figs. \ref{Fig:3DsketchPlusModes} and \ref{Fig:plasmonicDimerScatteredFields} it is evident that careful scattering calculations can indeed provide an approximation to the QNM field distribution at positions close to the resonator. Assuming a single mode expansion, one can then compare to an independent numerical calculation of the optical response at a single point in space to get the correct scaling of the field, in this way circumventing the need for a proper normalization~\cite{Bai_OE_21_27371_2013}. Given the dependence of the scattered field on the excitation condition as well as the possibility of additional QNMs at nearby resonance frequencies, it is clear that such an approach requires some care.}

\section*{{Mode volume and Purcell factor for leaky optical cavities}}
\rev{Modern formulations of light-matter interaction and scattering in general material systems are often based on the electromagnetic Green tensor~\cite{Tai_1994, Martin_PRE_58_3909_1998, NovotnyAndHecht_2006}. The Green tensor, which is closely related to the so-called local density of states (LDOS)~\cite{Sprik_EuroPhys_35_265_1996, NovotnyAndHecht_2006}, is the field propagator which may be interpreted as the field at $\mr$ due to a point source at $\mr'$. It is known analytically for certain simply geometries, but in general it must be calculated numerically; for example as the electric field response from a dipole source in either frequency-\cite{deLasson_JOSAB_30_1996_2013} 
or time-domain~\cite{YaoLaserReviews} scattering calculations. Common to all numerical calculations of the Green tensor, however, is the fact that they are rather expensive to compute for general material systems. For resonant systems, however, we expect most of the important physical processes of interest to be related to the resonant QNMs. Therefore, instead of full numerical solutions (which are often intractable), in such cases it may be both computationally and physically more appealing to formulate the light-matter interaction in terms of the QNMs, either directly or via the Green tensor. 
In Ref.~\onlinecite{Kristensen_PRB_88_205308_2013}, for example, a single-mode expansion of the Green tensor was used to dramatically simplify spontaneous emission calculations 
beyond the dipole approximation.}



Two of the most common and useful metrics for characterizing the \rev{properties of optical} cavities are the $Q$ value and the effective mode volume $V_\text{eff}$; a large $Q/V_\text{eff}$ ratio results in enhanced light-matter interactions \rev{as typically quantified by the LDOS. 
Physically, we can interpret the enhancement as being due to $(i)$ the light spending more time before leaking out of cavities with high $Q$ values, and $(ii)$ the electromagnetic field being enhanced when confined to small volumes}. This enhancement can be exploited in numerous photonic applications, including sensing, lasing, spasing, and quantum optics~\cite{Bergman_PRL_90_027402_2003,
Odom_NL_12_5769_2012, Maier_2007, Noginov_Nature_460_1110_2009, Novotny_NaturePhot_5_83_2011, Chang_NaturePhys_3_807_2007, Kusar_APL_100_231102_2012, Andersen_NaturePhys_7_215_2011, Tame_NaturePhys_9_329_2013,BeriniNatureLaser}
. The Purcell effect is a beautiful example of a situation in which a cavity with a large $Q/V_\text{eff}$ ratio enhances the spontaneous emission rate of an atom or QD. In general, the spontaneous emission rate $\Gamma_\alpha(\mr,\omega)$ of a dipole emitter with orientation $\me_\alpha$ may be enhanced or suppressed as compared to the rate $\Gamma_\text{B}$ in a homogeneous medium, and the enhancement is simply the relative LDOS~\cite{NovotnyAndHecht_2006}. In Purcell's original paper, a modest abstract published in the proceedings of the American Physical Society meeting at Cambridge in 1946~\cite{Purcell_PR_69_681_1946}, Purcell {formulated} the enhanced spontaneous emission factor {in a very elegant way} as
\begin{align}
F_\text{P} = \frac{3}{4\pi^2}\left(\frac{\lambda_\text{c}}{n_\text{c}}\right)^3\left(\frac{Q}{V_\text{eff}}\right),
\label{Eq:PurcellFactor}
\end{align}
where $\lambda_\text{c}$ is the free space wavelength, and $n_\text{c}$ is the refractive index at the dipole position. \rev{Purcell's formula assumes ($i$) the atom is at the field  maximum $\mr_\text{c}$ and resonant with the single cavity mode of interest, and ($ii$) the atom (or dipole emitter) has a dipole orientation that is the same as the polarization of the cavity mode.} Purcell was originally studying spontaneous emission rates at radio frequencies being enhanced in resonant electrical circuits, but the basic concepts of enhanced emission due to medium enhanced resonances applies to a wide range of frequencies. 

The mode volume $V_\text{eff}$ introduced by Purcell was essentially the physical volume of the resonator. It is customary to define the effective mode volume for \emph{normal modes}, $V^\text{N}_\text{eff}$, as
\begin{align}
V^\text{N}_\text{eff}=\rev{\frac{\langle\mf_\text{c}|\mf_\text{c}\rangle}{\epsilon(\mr_\text{c})|\mf_\text{c}(\mr_\text{c})|^2}} = \int_V \frac{\epsilon_\text{}(\mr) |\mf_\text{c}(\mr)|^2}{\epsilon(\mr_\text{c})|\mf_\text{c}(\mr_\text{c})|^2}  \ud\mr,
\label{Eq:VeffN}
\end{align}
where the integral is over all space.
This normal mode volume is a pure electromagnetic property and does not depend upon any embedded atoms or QDs. Physically, we can think of the mode volume as being a measure of the volume taken up by the normal mode. It is no exaggeration to say that Eq.~(\ref{Eq:VeffN}) has been the workhorse for cavity physics for decades, but it turns out to be wrong! At least it turns out to be wrong for any cavity with dissipation and hence a finite $Q$ value. The problem with Eq.~(\ref{Eq:VeffN}) when applied to leaky cavities is that it is based on the assumption that the mode is localized in space (or localized and periodic, such as for an optical waveguide mode). However, as we have already discussed, and {illustrated} explicitly in Fig.~\ref{Fig:plasmonicDimerModeDivergence}, all optical cavities have finite leakage 
which leads to modes that diverge in space. Consequently, the normal mode volume, as defined in (\ref{Eq:VeffN}), diverges exponentially \rev{when applied to the QNMs of cavities with a} finite $Q$ value if {taken at face value} and integrated over all space. 
For high-$Q$ cavities, the divergence as a function of integration volume is \rev{(initially)} rather slow~\cite{Kristensen_OL_37_1649_2012}, and the error in numerical calculations may in practice be small compared to errors or uncertainties in other theoretical parameters when performing the integration over typical calculation volumes. Nevertheless, the integral is in principle divergent, and one should instead use a generalized mode volume which is well defined and just as easy to calculate. Obviously, when dealing with low-$Q$ cavities or plasmonic nanoparticles this problem is much more severe and clearly a better approach is needed. 
By deriving the Purcell factor within a QNM picture (see Appendix~\ref{PurcellDerivation} for details), one can directly arrive at the Purcell factor in Eq.~(\ref{Eq:PurcellFactor}) with an effective mode volume $V_\text{eff}=V_\text{eff}^\text{Q}$ where
\begin{align}
\frac{1}{V_\text{eff}^\text{Q}} =  \text{Re}\left\{\frac{1}{v_\text{Q}}\right\},\quad v_\text{Q}=\frac{\langle\langle\mft_\text{c}|\mft_\text{c}\rangle\rangle}{\epsilon_\text{}(\mr_\text{c})\,\mft^2_\text{c}(\mr_\text{c})},
\label{Eq:VeffQ}
\end{align}
\rev{in which $\mft^2_\text{c}(\mr_\text{c}) = \mft_\text{c}(\mr_\text{c})\cdot\mft_\text{c}(\mr_\text{c})$}. This prescription provides a direct and unambiguous way of calculating the effective mode volume for leaky cavities, {including (dispersive) metal nanoparticle structures}, if a single mode approximation is valid (though extensions to include several modes is straightforward). 

\begin{figure}[t!]
\includegraphics[width=\columnwidth]{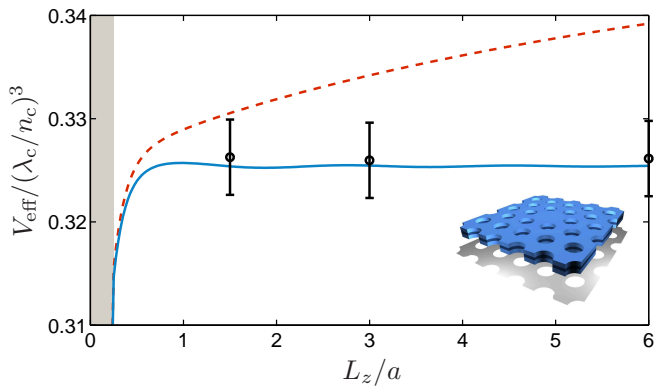}
%
\caption{\label{Fig:modeVolume_3D}Effective mode volume $V_{\rm eff}^\text{N}$ (red dashed) and $V_{\rm eff}^\text{Q}$ (blue solid) for the dielectric cavity in Fig. \ref{Fig:3DsketchPlusModes} as a function of height of the calculation domain $L_z$. Circles indicate reference mode volumes $V_\text{eff}^\text{tot}$ derived from independent Green tensor calculations \cite{YaoLaserReviews} with estimated error bars at different domain heights. Grey shaded area shows the extend of the membrane. {Figure from Ref.~\onlinecite{Kristensen_OL_37_1649_2012}.}
}
\end{figure}

For the photonic crystal cavity in Fig.~\ref{Fig:3DsketchPlusModes}, we show explicitly in Fig.~\ref{Fig:modeVolume_3D} that $V_\text{eff}^\text{N}$ diverges as a function of calculation domain size, whereas $V_\text{eff}^\text{Q}$ converges quickly to the correct value, as verified by rigorous numerical calculations. 
{Reference~\onlinecite{Kristensen_OL_37_1649_2012} compares the discrepancy between the two mode volumes for various $Q$ values, showing a dramatic divergence of the normal mode volume for low-$Q$ cavities.} 
\rev{
A similar rapid convergence of $V_\text{eff}^\text{Q}$ was found also in  Ref.~\onlinecite{oursMNP} for both two- and three-dimensional metallic nanorods, and shows that the 
divergent part of a QNM does not contribute to the effective mode volume. In this way we can make a connection to the work of Snyder and Love~\cite{SnyderLoveBook} who introduced a so-called ``caustic radius'' to describe the cross-over region, where leaky modes of optical waveguides no longer appear like bound waveguide modes. 
By use of the convergence analysis for the effective mode volume one can define a caustic radius for resonant systems also\cite{oursMNP}, and this caustic radius may then be interpreted in a physically appealing way as the boundary marking the extent of the QNM volume.

%


As noted above, Eq.~(\ref{Eq:PurcellFactor}) 
is based on the assumption that the emitter is at the field maximum (both spectrally and spatially) and that the dipole moment orientation $\me_\alpha$ is parallel to the field 
at this point. If this is not the case, then one can make a trivial generalization by multiplying by a factor $\eta(\mr,\me_\alpha,\omega)$ to account for any deviations; this is done, for example, in Ref.~\onlinecite{Gerard_JLT_17_2089_1999} for dielectric cavities. 
{In this way, the enhanced spontaneous emission rate at positions inside the cavity may be written in terms of the Purcell factor as 
\begin{align}
F_\alpha({\bf r},\omega)\equiv\frac{\Gamma_\alpha(\mr,\omega)}{\Gamma_\text{B}(\omega)} = F_\text{P}\eta(\mr,\me_\alpha,\omega).
\label{Eq:GeneralizedPurcell}
\end{align}
Moreover, there is an implicit assumption in the Purcell factor that the emitter couples to a single mode only. If this is not the case, then one can still derive the proper emission enhancement within the framework of QNMs by extending the methods in Refs.~\onlinecite{Kristensen_OL_37_1649_2012,Sauvan_PRL_110_237401_2013,oursMNP}} or the Appendix 
to include several modes.  Such an approach, however comes at the expense of the simplicity of Eq.~(\ref{Eq:PurcellFactor}).}



It is interesting to compare Eq.~(\ref{Eq:VeffQ}) to other definitions of an effective mode volume in the literature. Using a completely different approach based on the Lorentz reciprocity theorem, Sauvan \emph{et al.}\cite{Sauvan_PRL_110_237401_2013} recently derived an expression for the effective mode volume which (in the limit of non-magnetic materials) can be shown to be identical to Eq.~(\ref{Eq:VeffQ}). {This suggests that Eq.~(\ref{Eq:VeffQ}) is indeed the proper generalization of Eq.~(\ref{Eq:VeffN}) to leaky and dispersive cavities.} 
For metal resonators, $\epsilon(\omega)$ is complex which adds extra trouble to the use of Eq.~(\ref{Eq:VeffN}). 
In this case, the energy density has to be modified to account for loss and dispersion. To account for energy stored inside the metal resonator \rev{described by a Drude model, of the form $\epsilon(\omega) = \epsilon_\text{R}+\text{i}\epsilon_\text{I}=1-\omega_{\rm p}^2/(\omega^2+\text{i}\omega\gamma)$}, Maier introduced a modified effective mode volume for plasmonic systems~\cite{MaierOE}, essentially replacing the numerator in Eq.~(\ref{Eq:VeffN}) by \cite{Ruppin02} $(\epsilon_\text{R}(\mr)+2\omega\epsilon_\text{I}(\mr)/\gamma)|\mf_\text{c}(\mr)|^2$. This addresses the issue of a complex permittivity, but does not rectify the integration of spatially divergent modes. A related problem was 
discussed by Koenderink~\cite{Koenderink_OL_35_4208_2010}, who proposed to 
extract off the known linear divergence of the effective mode volume when using this modified energy density with computations using a scattered field solution. 


\section*{{Purcell factor and enhanced spontaneous emission rate for plasmonic systems}}


Although Eq.~(\ref{Eq:VeffQ}) provides an unambiguous and well-defined generalization of the effective mode volume to leaky and dispersive systems, this does not guarantee that the Purcell factor itself is a good approximation to the actual medium-enhanced spontaneous emission in general for plasmonic systems. \rev{A severe problem with the use of Eq.~(\ref{Eq:GeneralizedPurcell})} for plasmonic systems derives from the fact that close to any metal surface, the LDOS diverges as $1/z^3$, where $z$ is the distance from the dipole emitter to the metal surface. This divergence is caused by nonradiative decay and ohmic losses and means that the Purcell factor cannot be correct at short distances~\cite{Koenderink_OL_35_4208_2010}. In the opposite regime of long distances, the exponential divergence of the QNMs means that the Purcell factor cannot be correct either. {Indeed, in this limit the function $\eta(\mr,\mathbf{d},\omega)$ {diverges, clearly indicating the failure of a theory based on coupling to a single mode of the distant resonator}. \rev{Despite the problematic regimes at short and long distances, there  is an interesting intermediate region where one can still formulate the enhanced spontaneous emission in terms of the Purcell factor in a way similar to Eq.~(\ref{Eq:GeneralizedPurcell}), but with an additional factor to account for the background 
response~\cite{Sauvan_PRL_110_237401_2013,oursMNP}. 
Although a formulation in terms of the Purcell factor is in principle possible, it often more convenient to work with a formulation in terms of the electric field Green tensor~\cite{NovotnyAndHecht_2006,Martin_PRE_58_3909_1998} for general calculations of light emission and propagation in plasmonic systems. In many cases, however, one can benefit greatly from an expansion\cite{Lee_JOSAB_16_1409_1999} of the Green tensor on a single or a few QNMs of the plasmonic resonator as this may dramatically simplify the calculations. 
}


%

{As an illustrative example, we analyze the Purcell effect in the vicinity of a two-dimensional metallic nanorod, which supports a well defined dipole mode with a specific polarization \cite{oursMNP}.
A similar (three dimensional) example was recently given by Sauvan \emph{et al.} using a slightly different formulation of the inner product~\cite{Sauvan_PRL_110_237401_2013} (see also Ref.~\onlinecite{oursMNP}). For the metal, we assume a Drude model with $\omega_p = 1.26\times10^{16}~{\rm rad/s}$ and $\gamma=7\times10^{13}~{\rm rad/s}$. The rod has a width of $10$~nm and a length of $80$~nm and is located in a homogeneous space with refractive index $n_\text{B}=1.5$. We consider an emitter with dipole moment along the rod axis and located $10$~nm from the end facet of the rod. Figure~\ref{fig:PF} shows the near-field mode profile of the QNM \rev{as well as} 
the resulting 
\rev{enhanced spontaneous emission} factor as a function of frequency. 
Also, we show the results of independent and full numerical calculations of the relative LDOS, clearly illustrating the applicability of \rev{a single QNM approximation} 
to capture the full non-Lorentzian lineshape in this case. We refer to Yao \emph{et al.}~\cite{YaoLaserReviews} for details of LDOS calculations using 
FDTD~\cite{Lumerical}. Further details for this \rev{metal} nanorod calculation are given in Ref.~\onlinecite{oursMNP}, \rev{which also 
extends the QNM model to short and long distances}.}

\begin{figure}[htb]
\includegraphics[width=\columnwidth]{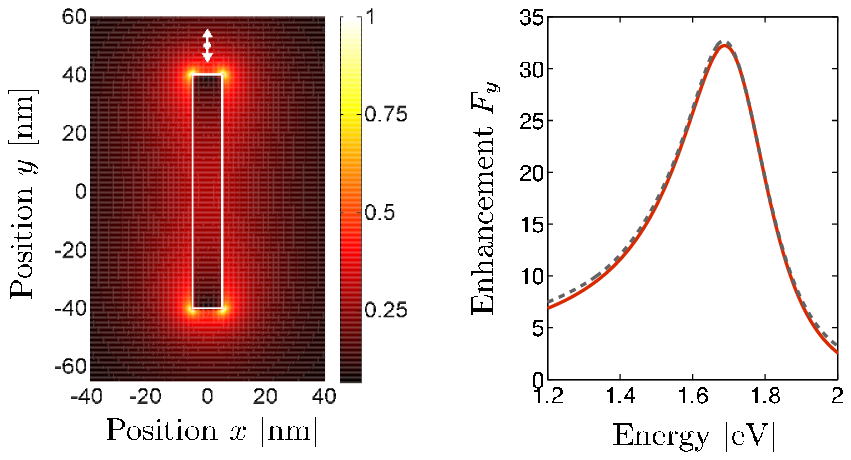}
%
\caption{\label{fig:PF}
(left) \rev{Near-field QNM mode} profile $|\tilde{\bf f}(x,y;\tilde\omega_{c})|$ for a \rev{metal nanorod}. (right) Enhanced spontaneous emission factor, $F_y({\bf r}_y,\omega)$,  at the location 10\,nm \rev{above} the end of a metal rod (see arrow in left figure); grey dashed is the \rev{full numerical solution 
and red solid is the single QNM approximation.} 
Figure adapted from Ref.~\onlinecite{oursMNP}.
}
\end{figure}

\section*{{Discussion and perspectives}}
It is certainly possible to describe optical phenomena such as spontaneous emission enhancement in 
cavities with a finite $Q$ value using more complicated and formal alternatives to Purcell's formula. One option is to calculate the LDOS from an expansion on normal modes, although a proper sampling of the continuum of modes for general cavities may be difficult in practice, or from so-called asymptotic in- and out-states which are built from coherent superpositions of solutions to the scattering problem~\cite{Liscidini_PRA_85_013833_2012}.
{In general, however, it not obvious that such an approach would 
be viable for three-dimensional problems with possibly lossy and dispersive cavities or nanoresonators of practical interest}. 
In spite of alternatives, Purcell's original formula and the concept of an effective mode volume is physically appealing and is very often \rev{the 
method of choice 
for calculating the expected emission enhancement.} 


In some contexts, cavity modes are presented merely as a physically appealing interpretation of spectral resonances that should be understood more correctly as coherent superpositions of normal modes. Although this is a valid interpretation, the QNM description shows that an equally valid interpretation is that of the cavity mode as a real physical quantity.  Each description has advantages and disadvantages depending on the particular physical problem, but neither should be considered more correct than the other. For the particular problem of the Purcell effect for leaky cavities or plasmonic nanoresonators, 
the QNM picture has the obvious advantage that only a single mode is needed, and Eq. (\ref{Eq:PurcellFactor}) applies with $V_\text{eff}=V_\text{eff}^\text{Q}$. One can {in principle} calculate the correct rate enhancement using a normal mode description, but this requires a continuum of modes. The distinction between normal modes and QNMs becomes important also in many hybrid systems of practical importance such as finite-sized waveguides~\cite{MangaRao_PRL_99_193901_2007} or coupled cavity-waveguide systems~\cite{Cowan_PRE_68_046606_2003,Yao_PRB_80_165128_2009}. In Ref.~\onlinecite{MangaRao_PRL_99_193901_2007}, the Purcell effect and effective mode volume was computed based on a numerical Green tensor approach~\cite{YaoLaserReviews}, although an alternative quasimode formulation would have been useful. 
For weak coupling between a waveguide and a cavity~\cite{Cowan_PRE_68_046606_2003, Yao_PRB_80_165128_2009}, a semi-analytical option is to use a rate equation approach to estimate the coupling between the modes in the cavity and the finite waveguide (both treated as normal modes). For strong waveguide-cavity coupling, however, this approach is not applicable. Again, for both weak and strong coupling, the QNM picture provides a direct and unambiguous alternative to the normal mode approach.

{
Last, we point towards a number of possible future applications of QNMs and pending problems for which the solutions would be of great \rev{value.} 
In many practical applications, one will be interested not only in the LDOS but also the electromagnetic propagators between different emitters~\cite{Hughes_PRL_94_227402_2005, Yao_OE_17_11505_2009, Kristensen_PRB_83_075305_2011} or from the emitter to the detector in the far field. In this case, too, it would be very useful to have a formulation in terms of the QNMs only. In particular, for the propagator to the far field, the divergence of the QNMs illustrates that this is a non-trivial problem.
\rev{A recent proposal\cite{oursMNP} offers a solution to this 
problem by use of a Dyson equation technique for the Green tensor and also introduces a simple and accurate way of including Ohmic losses at very short distances.} Another possible application is related to the so-called coupled mode theory~\cite{Pierce_JAP_25_179_1954, Haus_Proc_IEEE_79_1505_1991, Haus_1984, Joannopoulos2008} for coupled cavity waveguide systems, which implicitly seems to rely on a description of the cavity modes as QNMs\rev{\cite{Heuck_CLEO_2012_JW4A-6, Kristensen_APL_102_041107_2013}}. Nevertheless, a rigorous theoretical derivation of their coupling to the (normal) modes of the waveguides \rev{is} missing, and some authors seem to believe that such a theory does not make sense~\cite{Liscidini_PRA_85_013833_2012}. Finally, we note that an interesting possible extension of the theory of QNMs would be to include the effects of a non-local material response which has 
recently been attracting much attention~\cite{Ciraci_Science_337_1072_2012, Fernandez-Dom_PRL_108_106802_2012, Raza_OE_21_27344_2013}. 
The introduction of a non-local material response comes at the price of an extra material parameter describing the coherence length of the electron gas in the metal. Although this would suggest the introduction of an additional equation, it was recently shown that the full non-local response may be captured by a single wave equation~\cite{Toscano_NanoPhot_2_161_2013}.


\section*{Summary}
We have discussed the use of QNMs to describe resonant cavity modes for both leaky \rev{optical} cavities and plasmonic nanoresonators. The QNMs behave differently to the normal modes \rev{of most text books; most notably they have complex resonance frequencies and an exponential divergence at long distances.  Though fully expected for cavity systems with any finite $Q$ value, this attribute is typically ignored in most theoretical treatments which renders properties like the mode volume rather ambiguous. Nevertheless, the QNMs are exactly the same modes that are typically computed by the community so there is no added complexity in terms of computational electromagnetics associated with the formalism discussed in this article. The QNMs are physically appealing, intuitive, and can be used in efficient approximations to the electromagnetic Green tensor for use in a wide range of problems in classical and quantum optics and plasmonics using arbitrary lossy material systems\cite{GePRB2013, ColePRBStrongCoupling}. We have summarized some recent developments in the field including computational methods with associated potential pitfalls and the introduction of a generalized effective mode volume~\cite{Kristensen_OL_37_1649_2012}, and we have discussed how these concepts can be applied to both dielectric cavity structures and nanoplasmonic resonators.}


}

}

\acknowledgements
This work was supported by the Natural Sciences and  Engineering Research Council of Canada and the Danish Council for Independent Research (FTP 10-093651). We gratefully acknowledge Jeff Young, Cole Van Vlack, Rong-Chun Ge and Jakob Rosenkrantz de Lasson for discussions and for their contributions to 
Refs.~\onlinecite{Kristensen_OL_37_1649_2012, deLasson_JOSAB_30_1996_2013} and \onlinecite{oursMNP} whose graphs are shown in part in this  article.

\section*{Appendix: Purcell factor derivation}
\label{PurcellDerivation}
To derive the Purcell factor within the QNM picture we first introduce the \rev{electric field Green} tensor through~\cite{NovotnyAndHecht_2006,Martin_PRE_58_3909_1998}
\begin{equation}
\nabla\times\nabla\times\mG(\mr,\mr';\omega) - k_0^2\epsilon_\text{}(\mr)\mG(\mr,\mr';\omega) = \mathbf{I}\delta(\mr-\mr'),
\label{Eq:GWaveEquation}
\end{equation}
subject to the Silver-M\"uller Radiation condition. The Green tensor is the electromagnetic propagator and provides the proper framework for calculating light emission and scattering in general dielectric structures. In general, the
 relative emission rate may be expressed as~\cite{NovotnyAndHecht_2006}
\begin{align}
F_\alpha({\bf r},\omega)=\frac{\Gamma_\alpha(\mr,\omega)}{\Gamma_\text{B}(\omega)} = \frac{\text{Im}\left\{\me_\alpha\mG(\mr,\mr;\omega)\me_\alpha\right\}}{\text{Im}\left\{\me_\alpha\mG_\text{B}(\mr,\mr;\omega)\me_\alpha\right\}},
\label{Eq:PurcellFromLDOS}
\end{align}
where $\mG_\text{B}(\mr,\mr';\omega)$ is the Green tensor in a homogeneous medium with $\epsilon_\text{}(\mr)=\epsilon_\text{B}$ \cite{Martin_PRE_58_3909_1998}. \rev{For positions within the resonator, we expand the transverse part of the Green tensor as}~\cite{Lee_JOSAB_16_1409_1999}
\begin{align}
\mG^\text{T}(\mr,\mr';\omega) = c^2\sum_\mu\frac{\mft_\mu(\mr)\mft_\mu(\mr')}{2\tlo_\mu(\tlo_\mu-\omega)}.
\label{Eq:GreensTensorFromQuasiNormalModes}
\end{align}
The implicit assumption behind the notion of a cavity mode is that one term dominates the expansion of the Green tensor in Eq. (\ref{Eq:GreensTensorFromQuasiNormalModes}) and hence that the expansion can approximated by this term only. The Purcell factor may be viewed as the single mode limit of the relative decay rate in Eq. (\ref{Eq:PurcellFromLDOS}), evaluated at the field maximum $\mr_\text{c}$ and at the resonance frequency $\omega=\omega_\text{c}$. Starting from Eqs. (\ref{Eq:PurcellFromLDOS}) and (\ref{Eq:GreensTensorFromQuasiNormalModes}) with just a single term, and noting that $\text{Im}\{\mG(\mr,\mr;\omega)\} = \text{Im}\{\mG^\text{T}(\mr,\mr;\omega) \}$,  we have
\begin{align}
F_\text{P} &= \frac{6\pi c}{n_\text{c}\omega_\text{c}}\text{Im}\left\{\me_\text{c}\mG(\mr_\text{c},\mr_\text{c};\omega_\text{c})\me_\text{c}\right\} \nonumber\\
&=\frac{3\pi c^3}{n_\text{c}\omega_\text{c}}\text{Im}\left\{{\rm i}\frac{\mft_\text{c}^2(\mr_\text{c})}{\omega_\text{c}\,\gamma_\text{c}}\right\},
\end{align}
where we have discarded a small term $(\gamma_\text{c})^2$. We define
\begin{align}
\epsilon_\text{r}(\mr_\text{c})\mft_\text{c}^2(\mr_\text{c}) = \frac{\epsilon_\text{r}(\mr_\text{c})\mft^2_\text{c}(\mr_\text{c})}{\langle\langle\mft_\text{c}|\mft_\text{c}\rangle\rangle} \equiv \frac{1}{v_\text{Q}},
\end{align}
where $v_\text{Q}=v_\text{Q}^\text{R}+{\rm i}v_\text{Q}^\text{I}$. Using  $Q=\omega_\text{c}/2\gamma_\text{c}$ and $\epsilon_\text{}(\mr_\text{c})=n_\text{c}^2$, we can write the Purcell factor as in Eq.~(\ref{Eq:PurcellFactor}) with $V_\text{eff}$ as given in Eq.~(\ref{Eq:VeffQ}).
In the general case, where the emitter is spatially or spectrally detuned, or where the orientation of the dipole moment is different from the field, one can use a straightforward generalization of the above approach to write the (generalized) Purcell factor as in Eq.~(\ref{Eq:GeneralizedPurcell}).


\bibliography{bib1}

\begin{thebibliography}{92}%
\makeatletter
\providecommand \@ifxundefined [1]{%
 \@ifx{#1\undefined}
}%
\providecommand \@ifnum [1]{%
 \ifnum #1\expandafter \@firstoftwo
 \else \expandafter \@secondoftwo
 \fi
}%
\providecommand \@ifx [1]{%
 \ifx #1\expandafter \@firstoftwo
 \else \expandafter \@secondoftwo
 \fi
}%
\providecommand \natexlab [1]{#1}%
\providecommand \enquote  [1]{``#1''}%
\providecommand \bibnamefont  [1]{#1}%
\providecommand \bibfnamefont [1]{#1}%
\providecommand \citenamefont [1]{#1}%
\providecommand \href@noop [0]{\@secondoftwo}%
\providecommand \href [0]{\begingroup \@sanitize@url \@href}%
\providecommand \@href[1]{\@@startlink{#1}\@@href}%
\providecommand \@@href[1]{\endgroup#1\@@endlink}%
\providecommand \@sanitize@url [0]{\catcode `\\12\catcode `\$12\catcode
  `\&12\catcode `\#12\catcode `\^12\catcode `\_12\catcode `\%12\relax}%
\providecommand \@@startlink[1]{}%
\providecommand \@@endlink[0]{}%
\providecommand \url  [0]{\begingroup\@sanitize@url \@url }%
\providecommand \@url [1]{\endgroup\@href {#1}{\urlprefix }}%
\providecommand \urlprefix  [0]{URL }%
\providecommand \Eprint [0]{\href }%
\providecommand \doibase [0]{http://dx.doi.org/}%
\providecommand \selectlanguage [0]{\@gobble}%
\providecommand \bibinfo  [0]{\@secondoftwo}%
\providecommand \bibfield  [0]{\@secondoftwo}%
\providecommand \translation [1]{[#1]}%
\providecommand \BibitemOpen [0]{}%
\providecommand \bibitemStop [0]{}%
\providecommand \bibitemNoStop [0]{.\EOS\space}%
\providecommand \EOS [0]{\spacefactor3000\relax}%
\providecommand \BibitemShut  [1]{\csname bibitem#1\endcsname}%
\let\auto@bib@innerbib\@empty
\bibitem [{\citenamefont {Chang}\ and\ \citenamefont
  {Campillo}(1996)}]{ChangCampillo1996}%
  \BibitemOpen
  \bibfield  {author} {\bibinfo {author} {\bibfnamefont {R.~K.}\ \bibnamefont
  {Chang}}\ and\ \bibinfo {author} {\bibfnamefont {A.~J.}\ \bibnamefont
  {Campillo}},\ }\href@noop {} {\emph {\bibinfo {title} {Optical Processes in
  Microcavities}}}\ (\bibinfo  {publisher} {World Scientific},\ \bibinfo {year}
  {1996})\BibitemShut {NoStop}%
\bibitem [{\citenamefont {Reithmaier}\ \emph {et~al.}(2004)\citenamefont
  {Reithmaier}, \citenamefont {S\k{e}k}, \citenamefont {L\"offler},
  \citenamefont {Hofmann}, \citenamefont {Kuhn}, \citenamefont {Reitzenstein},
  \citenamefont {Keldysh}, \citenamefont {Kulakovskii}, \citenamefont
  {Reinecke},\ and\ \citenamefont {Forchel}}]{Reithmaier_Nature_42_197_2004}%
  \BibitemOpen
  \bibfield  {author} {\bibinfo {author} {\bibfnamefont {J.~P.}\ \bibnamefont
  {Reithmaier}}, \bibinfo {author} {\bibfnamefont {G.}~\bibnamefont {S\k{e}k}},
  \bibinfo {author} {\bibfnamefont {A.}~\bibnamefont {L\"offler}}, \bibinfo
  {author} {\bibfnamefont {C.}~\bibnamefont {Hofmann}}, \bibinfo {author}
  {\bibfnamefont {S.}~\bibnamefont {Kuhn}}, \bibinfo {author} {\bibfnamefont
  {S.}~\bibnamefont {Reitzenstein}}, \bibinfo {author} {\bibfnamefont {L.~V.}\
  \bibnamefont {Keldysh}}, \bibinfo {author} {\bibfnamefont {V.~D.}\
  \bibnamefont {Kulakovskii}}, \bibinfo {author} {\bibfnamefont {T.~L.}\
  \bibnamefont {Reinecke}}, \ and\ \bibinfo {author} {\bibfnamefont
  {A.}~\bibnamefont {Forchel}},\ }\href@noop {} {\bibfield  {journal} {\bibinfo
   {journal} {Nature}\ }\textbf {\bibinfo {volume} {432}},\ \bibinfo {pages}
  {197} (\bibinfo {year} {2004})}\BibitemShut {NoStop}%
\bibitem [{\citenamefont {Yoshie}\ \emph {et~al.}(2004)\citenamefont {Yoshie},
  \citenamefont {Scherer}, \citenamefont {Hendrickson}, \citenamefont
  {Khitrova}, \citenamefont {Gibbs}, \citenamefont {Rupper}, \citenamefont
  {Ell}, \citenamefont {Shchekin},\ and\ \citenamefont
  {Deppe}}]{Yoshie_Nature_432_200_2004}%
  \BibitemOpen
  \bibfield  {author} {\bibinfo {author} {\bibfnamefont {T.}~\bibnamefont
  {Yoshie}}, \bibinfo {author} {\bibfnamefont {A.}~\bibnamefont {Scherer}},
  \bibinfo {author} {\bibfnamefont {J.}~\bibnamefont {Hendrickson}}, \bibinfo
  {author} {\bibfnamefont {G.}~\bibnamefont {Khitrova}}, \bibinfo {author}
  {\bibfnamefont {H.~M.}\ \bibnamefont {Gibbs}}, \bibinfo {author}
  {\bibfnamefont {G.}~\bibnamefont {Rupper}}, \bibinfo {author} {\bibfnamefont
  {C.}~\bibnamefont {Ell}}, \bibinfo {author} {\bibfnamefont {O.~B.}\
  \bibnamefont {Shchekin}}, \ and\ \bibinfo {author} {\bibfnamefont {D.~G.}\
  \bibnamefont {Deppe}},\ }\href@noop {} {\bibfield  {journal} {\bibinfo
  {journal} {Nature}\ }\textbf {\bibinfo {volume} {432}},\ \bibinfo {pages}
  {200} (\bibinfo {year} {2004})}\BibitemShut {NoStop}%
\bibitem [{\citenamefont {Press}\ \emph {et~al.}(2007)\citenamefont {Press},
  \citenamefont {G\"otzinger}, \citenamefont {Reitzenstein}, \citenamefont
  {Hofmann}, \citenamefont {L\"offler}, \citenamefont {Kamp}, \citenamefont
  {Forchel},\ and\ \citenamefont {Yamamoto}}]{Press_PRL_98_117402_2007}%
  \BibitemOpen
  \bibfield  {author} {\bibinfo {author} {\bibfnamefont {D.}~\bibnamefont
  {Press}}, \bibinfo {author} {\bibfnamefont {S.}~\bibnamefont {G\"otzinger}},
  \bibinfo {author} {\bibfnamefont {S.}~\bibnamefont {Reitzenstein}}, \bibinfo
  {author} {\bibfnamefont {C.}~\bibnamefont {Hofmann}}, \bibinfo {author}
  {\bibfnamefont {A.}~\bibnamefont {L\"offler}}, \bibinfo {author}
  {\bibfnamefont {M.}~\bibnamefont {Kamp}}, \bibinfo {author} {\bibfnamefont
  {A.}~\bibnamefont {Forchel}}, \ and\ \bibinfo {author} {\bibfnamefont
  {Y.}~\bibnamefont {Yamamoto}},\ }\href@noop {} {\bibfield  {journal}
  {\bibinfo  {journal} {Physical Review Letters}\ }\textbf {\bibinfo {volume}
  {98}},\ \bibinfo {pages} {117402} (\bibinfo {year} {2007})}\BibitemShut
  {NoStop}%
\bibitem [{\citenamefont {Vahala}(2003)}]{Vahala_Nature_424_839_2003}%
  \BibitemOpen
  \bibfield  {author} {\bibinfo {author} {\bibfnamefont {K.~J.}\ \bibnamefont
  {Vahala}},\ }\href@noop {} {\bibfield  {journal} {\bibinfo  {journal}
  {Nature}\ }\textbf {\bibinfo {volume} {424}},\ \bibinfo {pages} {839}
  (\bibinfo {year} {2003})}\BibitemShut {NoStop}%
\bibitem [{\citenamefont {Bergman}\ and\ \citenamefont
  {Stockman}(2003)}]{Bergman_PRL_90_027402_2003}%
  \BibitemOpen
  \bibfield  {author} {\bibinfo {author} {\bibfnamefont {D.~J.}\ \bibnamefont
  {Bergman}}\ and\ \bibinfo {author} {\bibfnamefont {M.~I.}\ \bibnamefont
  {Stockman}},\ }\href@noop {} {\bibfield  {journal} {\bibinfo  {journal}
  {Physical Review Letters}\ }\textbf {\bibinfo {volume} {90}},\ \bibinfo
  {pages} {027402} (\bibinfo {year} {2003})}\BibitemShut {NoStop}%
\bibitem [{\citenamefont {Suh}\ \emph {et~al.}(2012)\citenamefont {Suh},
  \citenamefont {Kim}, \citenamefont {Zhou}, \citenamefont {Huntington},
  \citenamefont {Co}, \citenamefont {Wasielewski},\ and\ \citenamefont
  {Odom}}]{Odom_NL_12_5769_2012}%
  \BibitemOpen
  \bibfield  {author} {\bibinfo {author} {\bibfnamefont {J.~Y.}\ \bibnamefont
  {Suh}}, \bibinfo {author} {\bibfnamefont {C.~H.}\ \bibnamefont {Kim}},
  \bibinfo {author} {\bibfnamefont {W.}~\bibnamefont {Zhou}}, \bibinfo {author}
  {\bibfnamefont {M.~D.}\ \bibnamefont {Huntington}}, \bibinfo {author}
  {\bibfnamefont {D.~T.}\ \bibnamefont {Co}}, \bibinfo {author} {\bibfnamefont
  {M.~R.}\ \bibnamefont {Wasielewski}}, \ and\ \bibinfo {author} {\bibfnamefont
  {T.~W.}\ \bibnamefont {Odom}},\ }\href@noop {} {\bibfield  {journal}
  {\bibinfo  {journal} {Nano Letters}\ }\textbf {\bibinfo {volume} {12}},\
  \bibinfo {pages} {5769} (\bibinfo {year} {2012})}\BibitemShut {NoStop}%
\bibitem [{\citenamefont {Maier}(2007)}]{Maier_2007}%
  \BibitemOpen
  \bibfield  {author} {\bibinfo {author} {\bibfnamefont {S.~A.}\ \bibnamefont
  {Maier}},\ }\href@noop {} {\emph {\bibinfo {title} {Plasmonics: Fundamentals
  and Applications}}}\ (\bibinfo  {publisher} {Springer},\ \bibinfo {year}
  {2007})\BibitemShut {NoStop}%
\bibitem [{\citenamefont {Noginov}\ \emph {et~al.}(2009)\citenamefont
  {Noginov}, \citenamefont {Zhu}, \citenamefont {Belgrave}, \citenamefont
  {Bakker}, \citenamefont {Shalaev}, \citenamefont {Narimanov}, \citenamefont
  {Stout}, \citenamefont {Herz}, \citenamefont {Suteewong},\ and\ \citenamefont
  {Wiesner}}]{Noginov_Nature_460_1110_2009}%
  \BibitemOpen
  \bibfield  {author} {\bibinfo {author} {\bibfnamefont {M.~A.}\ \bibnamefont
  {Noginov}}, \bibinfo {author} {\bibfnamefont {G.}~\bibnamefont {Zhu}},
  \bibinfo {author} {\bibfnamefont {A.~M.}\ \bibnamefont {Belgrave}}, \bibinfo
  {author} {\bibfnamefont {R.}~\bibnamefont {Bakker}}, \bibinfo {author}
  {\bibfnamefont {V.~M.}\ \bibnamefont {Shalaev}}, \bibinfo {author}
  {\bibfnamefont {E.~E.}\ \bibnamefont {Narimanov}}, \bibinfo {author}
  {\bibfnamefont {S.}~\bibnamefont {Stout}}, \bibinfo {author} {\bibfnamefont
  {E.}~\bibnamefont {Herz}}, \bibinfo {author} {\bibfnamefont {T.}~\bibnamefont
  {Suteewong}}, \ and\ \bibinfo {author} {\bibfnamefont {U.}~\bibnamefont
  {Wiesner}},\ }\href@noop {} {\bibfield  {journal} {\bibinfo  {journal}
  {Nature}\ }\textbf {\bibinfo {volume} {460}},\ \bibinfo {pages} {1110}
  (\bibinfo {year} {2009})}\BibitemShut {NoStop}%
\bibitem [{\citenamefont {Novotny}\ and\ \citenamefont {van
  Hulst}(2011)}]{Novotny_NaturePhot_5_83_2011}%
  \BibitemOpen
  \bibfield  {author} {\bibinfo {author} {\bibfnamefont {L.}~\bibnamefont
  {Novotny}}\ and\ \bibinfo {author} {\bibfnamefont {N.}~\bibnamefont {van
  Hulst}},\ }\href@noop {} {\bibfield  {journal} {\bibinfo  {journal} {Nature
  Photonics}\ }\textbf {\bibinfo {volume} {5}},\ \bibinfo {pages} {83}
  (\bibinfo {year} {2011})}\BibitemShut {NoStop}%
\bibitem [{\citenamefont {Chang}\ \emph {et~al.}(2007)\citenamefont {Chang},
  \citenamefont {S{\o}rensen}, \citenamefont {Demler},\ and\ \citenamefont
  {Lukin}}]{Chang_NaturePhys_3_807_2007}%
  \BibitemOpen
  \bibfield  {author} {\bibinfo {author} {\bibfnamefont {D.~E.}\ \bibnamefont
  {Chang}}, \bibinfo {author} {\bibfnamefont {A.~S.}\ \bibnamefont
  {S{\o}rensen}}, \bibinfo {author} {\bibfnamefont {E.~A.}\ \bibnamefont
  {Demler}}, \ and\ \bibinfo {author} {\bibfnamefont {M.~D.}\ \bibnamefont
  {Lukin}},\ }\href@noop {} {\bibfield  {journal} {\bibinfo  {journal} {Nature
  Physics}\ }\textbf {\bibinfo {volume} {3}},\ \bibinfo {pages} {807} (\bibinfo
  {year} {2007})}\BibitemShut {NoStop}%
\bibitem [{\citenamefont {Gruber}\ \emph {et~al.}(2012)\citenamefont {Gruber},
  \citenamefont {Kusar}, \citenamefont {Hohenau},\ and\ \citenamefont
  {Krenn}}]{Kusar_APL_100_231102_2012}%
  \BibitemOpen
  \bibfield  {author} {\bibinfo {author} {\bibfnamefont {C.}~\bibnamefont
  {Gruber}}, \bibinfo {author} {\bibfnamefont {P.}~\bibnamefont {Kusar}},
  \bibinfo {author} {\bibfnamefont {A.}~\bibnamefont {Hohenau}}, \ and\
  \bibinfo {author} {\bibfnamefont {J.~R.}\ \bibnamefont {Krenn}},\ }\href@noop
  {} {\bibfield  {journal} {\bibinfo  {journal} {Applied Physics Letters}\
  }\textbf {\bibinfo {volume} {100}},\ \bibinfo {pages} {231102} (\bibinfo
  {year} {2012})}\BibitemShut {NoStop}%
\bibitem [{\citenamefont {Andersen}\ \emph {et~al.}(2011)\citenamefont
  {Andersen}, \citenamefont {Stobbe}, \citenamefont {S{\o}rensen},\ and\
  \citenamefont {Lodahl}}]{Andersen_NaturePhys_7_215_2011}%
  \BibitemOpen
  \bibfield  {author} {\bibinfo {author} {\bibfnamefont {M.~L.}\ \bibnamefont
  {Andersen}}, \bibinfo {author} {\bibfnamefont {S.}~\bibnamefont {Stobbe}},
  \bibinfo {author} {\bibfnamefont {A.~S.}\ \bibnamefont {S{\o}rensen}}, \ and\
  \bibinfo {author} {\bibfnamefont {P.}~\bibnamefont {Lodahl}},\ }\href@noop {}
  {\bibfield  {journal} {\bibinfo  {journal} {Nature Physics}\ }\textbf
  {\bibinfo {volume} {7}},\ \bibinfo {pages} {215} (\bibinfo {year}
  {2011})}\BibitemShut {NoStop}%
\bibitem [{\citenamefont {Tame}\ \emph {et~al.}(2013)\citenamefont {Tame},
  \citenamefont {McEnery}, \citenamefont {\"{O}zdemir}, \citenamefont {Lee},
  \citenamefont {Maier},\ and\ \citenamefont
  {Kim}}]{Tame_NaturePhys_9_329_2013}%
  \BibitemOpen
  \bibfield  {author} {\bibinfo {author} {\bibfnamefont {M.~S.}\ \bibnamefont
  {Tame}}, \bibinfo {author} {\bibfnamefont {K.~R.}\ \bibnamefont {McEnery}},
  \bibinfo {author} {\bibfnamefont {S.~K.}\ \bibnamefont {\"{O}zdemir}},
  \bibinfo {author} {\bibfnamefont {J.}~\bibnamefont {Lee}}, \bibinfo {author}
  {\bibfnamefont {S.~A.}\ \bibnamefont {Maier}}, \ and\ \bibinfo {author}
  {\bibfnamefont {M.~S.}\ \bibnamefont {Kim}},\ }\href@noop {} {\bibfield
  {journal} {\bibinfo  {journal} {Nature Physics}\ }\textbf {\bibinfo {volume}
  {9}},\ \bibinfo {pages} {329} (\bibinfo {year} {2013})}\BibitemShut {NoStop}%
\bibitem [{\citenamefont {Berini}\ and\ \citenamefont
  {Leon}(2012)}]{BeriniNatureLaser}%
  \BibitemOpen
  \bibfield  {author} {\bibinfo {author} {\bibfnamefont {P.}~\bibnamefont
  {Berini}}\ and\ \bibinfo {author} {\bibfnamefont {I.~D.}\ \bibnamefont
  {Leon}},\ }\href@noop {} {\bibfield  {journal} {\bibinfo  {journal} {Nature
  Photonics}\ }\textbf {\bibinfo {volume} {6}},\ \bibinfo {pages} {16}
  (\bibinfo {year} {2012})}\BibitemShut {NoStop}%
\bibitem [{\citenamefont {Belacel}\ \emph {et~al.}(2013)\citenamefont
  {Belacel}, \citenamefont {Habert}, \citenamefont {Bigourdan}, \citenamefont
  {Marquier}, \citenamefont {Hugonin}, \citenamefont {de~Vasconcellos},
  \citenamefont {Lafosse}, \citenamefont {Coolen}, \citenamefont {Schwob},
  \citenamefont {Javaux}, \citenamefont {Dubertret}, \citenamefont {Greffet},
  \citenamefont {Senellart},\ and\ \citenamefont
  {Maitre}}]{Belacel_NL_13_1516_2013}%
  \BibitemOpen
  \bibfield  {author} {\bibinfo {author} {\bibfnamefont {C.}~\bibnamefont
  {Belacel}}, \bibinfo {author} {\bibfnamefont {B.}~\bibnamefont {Habert}},
  \bibinfo {author} {\bibfnamefont {F.}~\bibnamefont {Bigourdan}}, \bibinfo
  {author} {\bibfnamefont {F.}~\bibnamefont {Marquier}}, \bibinfo {author}
  {\bibfnamefont {J.-P.}\ \bibnamefont {Hugonin}}, \bibinfo {author}
  {\bibfnamefont {S.~M.}\ \bibnamefont {de~Vasconcellos}}, \bibinfo {author}
  {\bibfnamefont {X.}~\bibnamefont {Lafosse}}, \bibinfo {author} {\bibfnamefont
  {L.}~\bibnamefont {Coolen}}, \bibinfo {author} {\bibfnamefont
  {C.}~\bibnamefont {Schwob}}, \bibinfo {author} {\bibfnamefont
  {C.}~\bibnamefont {Javaux}}, \bibinfo {author} {\bibfnamefont
  {B.}~\bibnamefont {Dubertret}}, \bibinfo {author} {\bibfnamefont {J.-J.}\
  \bibnamefont {Greffet}}, \bibinfo {author} {\bibfnamefont {P.}~\bibnamefont
  {Senellart}}, \ and\ \bibinfo {author} {\bibfnamefont {A.}~\bibnamefont
  {Maitre}},\ }\href@noop {} {\bibfield  {journal} {\bibinfo  {journal} {Nano
  Letters}\ }\textbf {\bibinfo {volume} {13}},\ \bibinfo {pages} {1516}
  (\bibinfo {year} {2013})}\BibitemShut {NoStop}%
\bibitem [{\citenamefont {Anetsberger}\ \emph {et~al.}(2008)\citenamefont
  {Anetsberger}, \citenamefont {Rivi\'ere}, \citenamefont {Schliesser},
  \citenamefont {Arcizet},\ and\ \citenamefont
  {Kippenberg}}]{AnetsbergerOptoMech}%
  \BibitemOpen
  \bibfield  {author} {\bibinfo {author} {\bibfnamefont {G.}~\bibnamefont
  {Anetsberger}}, \bibinfo {author} {\bibfnamefont {R.}~\bibnamefont
  {Rivi\'ere}}, \bibinfo {author} {\bibfnamefont {A.}~\bibnamefont
  {Schliesser}}, \bibinfo {author} {\bibfnamefont {O.}~\bibnamefont {Arcizet}},
  \ and\ \bibinfo {author} {\bibfnamefont {T.~J.}\ \bibnamefont {Kippenberg}},\
  }\href@noop {} {\bibfield  {journal} {\bibinfo  {journal} {Nature Photonics}\
  }\textbf {\bibinfo {volume} {2}},\ \bibinfo {pages} {627} (\bibinfo {year}
  {2008})}\BibitemShut {NoStop}%
\bibitem [{\citenamefont {Thompson}\ \emph {et~al.}(2008)\citenamefont
  {Thompson}, \citenamefont {Zwickl}, \citenamefont {Jayich}, \citenamefont
  {Marquardt}, \citenamefont {Girvin},\ and\ \citenamefont
  {Harris}}]{ThompsonOptMech}%
  \BibitemOpen
  \bibfield  {author} {\bibinfo {author} {\bibfnamefont {J.~D.}\ \bibnamefont
  {Thompson}}, \bibinfo {author} {\bibfnamefont {B.~M.}\ \bibnamefont
  {Zwickl}}, \bibinfo {author} {\bibfnamefont {A.~M.}\ \bibnamefont {Jayich}},
  \bibinfo {author} {\bibfnamefont {F.}~\bibnamefont {Marquardt}}, \bibinfo
  {author} {\bibfnamefont {S.~M.}\ \bibnamefont {Girvin}}, \ and\ \bibinfo
  {author} {\bibfnamefont {J.~G.~E.}\ \bibnamefont {Harris}},\ }\href@noop {}
  {\bibfield  {journal} {\bibinfo  {journal} {Nature}\ }\textbf {\bibinfo
  {volume} {452}},\ \bibinfo {pages} {72} (\bibinfo {year} {2008})}\BibitemShut
  {NoStop}%
\bibitem [{\citenamefont {Safavi-Naeini}\ \emph {et~al.}(2012)\citenamefont
  {Safavi-Naeini}, \citenamefont {Chan}, \citenamefont {Hill}, \citenamefont
  {Alegre}, \citenamefont {Krause},\ and\ \citenamefont
  {Painter}}]{PainterOptoMech1}%
  \BibitemOpen
  \bibfield  {author} {\bibinfo {author} {\bibfnamefont {A.~H.}\ \bibnamefont
  {Safavi-Naeini}}, \bibinfo {author} {\bibfnamefont {J.}~\bibnamefont {Chan}},
  \bibinfo {author} {\bibfnamefont {J.~T.}\ \bibnamefont {Hill}}, \bibinfo
  {author} {\bibfnamefont {T.~P.~M.}\ \bibnamefont {Alegre}}, \bibinfo {author}
  {\bibfnamefont {A.}~\bibnamefont {Krause}}, \ and\ \bibinfo {author}
  {\bibfnamefont {O.}~\bibnamefont {Painter}},\ }\href@noop {} {\bibfield
  {journal} {\bibinfo  {journal} {Physical Review Letters}\ }\textbf {\bibinfo
  {volume} {108}},\ \bibinfo {pages} {033602} (\bibinfo {year}
  {2012})}\BibitemShut {NoStop}%
\bibitem [{\citenamefont {Kippenberg}\ and\ \citenamefont
  {Vahala}(2008)}]{KippenbergOptoMech1}%
  \BibitemOpen
  \bibfield  {author} {\bibinfo {author} {\bibfnamefont {T.~J.}\ \bibnamefont
  {Kippenberg}}\ and\ \bibinfo {author} {\bibfnamefont {K.~J.}\ \bibnamefont
  {Vahala}},\ }\href@noop {} {\bibfield  {journal} {\bibinfo  {journal}
  {Science}\ }\textbf {\bibinfo {volume} {321}},\ \bibinfo {pages} {1172}
  (\bibinfo {year} {2008})}\BibitemShut {NoStop}%
\bibitem [{\citenamefont {Ge}\ \emph {et~al.}(2013{\natexlab{a}})\citenamefont
  {Ge}, \citenamefont {Vlack}, \citenamefont {Yao}, \citenamefont {Young},\
  and\ \citenamefont {Hughes}}]{GePRB2013}%
  \BibitemOpen
  \bibfield  {author} {\bibinfo {author} {\bibfnamefont {R.-C.}\ \bibnamefont
  {Ge}}, \bibinfo {author} {\bibfnamefont {C.~V.}\ \bibnamefont {Vlack}},
  \bibinfo {author} {\bibfnamefont {P.}~\bibnamefont {Yao}}, \bibinfo {author}
  {\bibfnamefont {J.~F.}\ \bibnamefont {Young}}, \ and\ \bibinfo {author}
  {\bibfnamefont {S.}~\bibnamefont {Hughes}},\ }\href@noop {} {\bibfield
  {journal} {\bibinfo  {journal} {Physical Review B}\ }\textbf {\bibinfo
  {volume} {87}},\ \bibinfo {pages} {205425} (\bibinfo {year}
  {2013}{\natexlab{a}})}\BibitemShut {NoStop}%
\bibitem [{\citenamefont {Tr\"ugler}\ and\ \citenamefont
  {Hohenester}(2008)}]{HohenesterPRB2008}%
  \BibitemOpen
  \bibfield  {author} {\bibinfo {author} {\bibfnamefont {A.}~\bibnamefont
  {Tr\"ugler}}\ and\ \bibinfo {author} {\bibfnamefont {U.}~\bibnamefont
  {Hohenester}},\ }\href@noop {} {\bibfield  {journal} {\bibinfo  {journal}
  {Physical Review B}\ }\textbf {\bibinfo {volume} {77}},\ \bibinfo {pages}
  {115403} (\bibinfo {year} {2008})}\BibitemShut {NoStop}%
\bibitem [{\citenamefont {Savasta}\ \emph {et~al.}(2010)\citenamefont
  {Savasta}, \citenamefont {Saija}, \citenamefont {Ridolfo}, \citenamefont
  {Stefano}, \citenamefont {Denti},\ and\ \citenamefont
  {Borghese}}]{SavastaNano}%
  \BibitemOpen
  \bibfield  {author} {\bibinfo {author} {\bibfnamefont {S.}~\bibnamefont
  {Savasta}}, \bibinfo {author} {\bibfnamefont {R.}~\bibnamefont {Saija}},
  \bibinfo {author} {\bibfnamefont {A.}~\bibnamefont {Ridolfo}}, \bibinfo
  {author} {\bibfnamefont {O.~D.}\ \bibnamefont {Stefano}}, \bibinfo {author}
  {\bibfnamefont {P.}~\bibnamefont {Denti}}, \ and\ \bibinfo {author}
  {\bibfnamefont {F.}~\bibnamefont {Borghese}},\ }\href@noop {} {\bibfield
  {journal} {\bibinfo  {journal} {ACS Nano}\ }\textbf {\bibinfo {volume} {4}},\
  \bibinfo {pages} {6369} (\bibinfo {year} {2010})}\BibitemShut {NoStop}%
\bibitem [{\citenamefont {Vlack}\ \emph {et~al.}(2012)\citenamefont {Vlack},
  \citenamefont {Kristensen},\ and\ \citenamefont
  {Hughes}}]{ColePRBStrongCoupling}%
  \BibitemOpen
  \bibfield  {author} {\bibinfo {author} {\bibfnamefont {C.~V.}\ \bibnamefont
  {Vlack}}, \bibinfo {author} {\bibfnamefont {P.~T.}\ \bibnamefont
  {Kristensen}}, \ and\ \bibinfo {author} {\bibfnamefont {S.}~\bibnamefont
  {Hughes}},\ }\href@noop {} {\bibfield  {journal} {\bibinfo  {journal}
  {Physical Review B}\ }\textbf {\bibinfo {volume} {85}},\ \bibinfo {pages}
  {075303} (\bibinfo {year} {2012})}\BibitemShut {NoStop}%
\bibitem [{\citenamefont {Kristensen}\ \emph {et~al.}(2012)\citenamefont
  {Kristensen}, \citenamefont {Vlack},\ and\ \citenamefont
  {Hughes}}]{Kristensen_OL_37_1649_2012}%
  \BibitemOpen
  \bibfield  {author} {\bibinfo {author} {\bibfnamefont {P.~T.}\ \bibnamefont
  {Kristensen}}, \bibinfo {author} {\bibfnamefont {C.~V.}\ \bibnamefont
  {Vlack}}, \ and\ \bibinfo {author} {\bibfnamefont {S.}~\bibnamefont
  {Hughes}},\ }\href@noop {} {\bibfield  {journal} {\bibinfo  {journal} {Optics
  Letters}\ }\textbf {\bibinfo {volume} {37}},\ \bibinfo {pages} {1649}
  (\bibinfo {year} {2012})}\BibitemShut {NoStop}%
\bibitem [{\citenamefont {de~Lasson}\ \emph {et~al.}(2013)\citenamefont
  {de~Lasson}, \citenamefont {M{\o}rk},\ and\ \citenamefont
  {Kristensen}}]{deLasson_JOSAB_30_1996_2013}%
  \BibitemOpen
  \bibfield  {author} {\bibinfo {author} {\bibfnamefont {J.~R.}\ \bibnamefont
  {de~Lasson}}, \bibinfo {author} {\bibfnamefont {J.}~\bibnamefont {M{\o}rk}},
  \ and\ \bibinfo {author} {\bibfnamefont {P.~T.}\ \bibnamefont {Kristensen}},\
  }\href@noop {} {\bibfield  {journal} {\bibinfo  {journal} {Journal of the
  Optical Society of America B}\ }\textbf {\bibinfo {volume} {30}},\ \bibinfo
  {pages} {1996} (\bibinfo {year} {2013})}\BibitemShut {NoStop}%
\bibitem [{\citenamefont {Barth}\ \emph {et~al.}(2010)\citenamefont {Barth},
  \citenamefont {Schietinger}, \citenamefont {Fischer}, \citenamefont {Becker},
  \citenamefont {N\"usse}, \citenamefont {Aichele}, \citenamefont {L\"ochel},
  \citenamefont {S\"onnichsen},\ and\ \citenamefont {Benson}}]{Barth}%
  \BibitemOpen
  \bibfield  {author} {\bibinfo {author} {\bibfnamefont {M.}~\bibnamefont
  {Barth}}, \bibinfo {author} {\bibfnamefont {S.}~\bibnamefont {Schietinger}},
  \bibinfo {author} {\bibfnamefont {S.}~\bibnamefont {Fischer}}, \bibinfo
  {author} {\bibfnamefont {J.}~\bibnamefont {Becker}}, \bibinfo {author}
  {\bibfnamefont {N.}~\bibnamefont {N\"usse}}, \bibinfo {author} {\bibfnamefont
  {T.}~\bibnamefont {Aichele}}, \bibinfo {author} {\bibfnamefont
  {B.}~\bibnamefont {L\"ochel}}, \bibinfo {author} {\bibfnamefont
  {C.}~\bibnamefont {S\"onnichsen}}, \ and\ \bibinfo {author} {\bibfnamefont
  {O.}~\bibnamefont {Benson}},\ }\href@noop {} {\bibfield  {journal} {\bibinfo
  {journal} {Nano Letters}\ }\textbf {\bibinfo {volume} {10}},\ \bibinfo
  {pages} {891} (\bibinfo {year} {2010})}\BibitemShut {NoStop}%
\bibitem [{\citenamefont {Mukherjee}\ \emph {et~al.}(2011)\citenamefont
  {Mukherjee}, \citenamefont {Hajisalem},\ and\ \citenamefont
  {Gordon}}]{Gordon12}%
  \BibitemOpen
  \bibfield  {author} {\bibinfo {author} {\bibfnamefont {I.}~\bibnamefont
  {Mukherjee}}, \bibinfo {author} {\bibfnamefont {G.}~\bibnamefont
  {Hajisalem}}, \ and\ \bibinfo {author} {\bibfnamefont {R.}~\bibnamefont
  {Gordon}},\ }\href@noop {} {\bibfield  {journal} {\bibinfo  {journal} {Optics
  Express}\ }\textbf {\bibinfo {volume} {19}},\ \bibinfo {pages} {22462}
  (\bibinfo {year} {2011})}\BibitemShut {NoStop}%
\bibitem [{\citenamefont {Lai}\ \emph {et~al.}(1990)\citenamefont {Lai},
  \citenamefont {Leung}, \citenamefont {Young}, \citenamefont {Barber},\ and\
  \citenamefont {Hill}}]{Lai_PRA_41_5187_1990}%
  \BibitemOpen
  \bibfield  {author} {\bibinfo {author} {\bibfnamefont {H.~M.}\ \bibnamefont
  {Lai}}, \bibinfo {author} {\bibfnamefont {P.~T.}\ \bibnamefont {Leung}},
  \bibinfo {author} {\bibfnamefont {K.}~\bibnamefont {Young}}, \bibinfo
  {author} {\bibfnamefont {P.~W.}\ \bibnamefont {Barber}}, \ and\ \bibinfo
  {author} {\bibfnamefont {S.~C.}\ \bibnamefont {Hill}},\ }\href@noop {}
  {\bibfield  {journal} {\bibinfo  {journal} {Physical Review A}\ }\textbf
  {\bibinfo {volume} {41}},\ \bibinfo {pages} {5187} (\bibinfo {year}
  {1990})}\BibitemShut {NoStop}%
\bibitem [{\citenamefont {Leung}\ \emph
  {et~al.}(1994{\natexlab{a}})\citenamefont {Leung}, \citenamefont {Liu},\ and\
  \citenamefont {Young}}]{Leung_PRA_49_3057_1994}%
  \BibitemOpen
  \bibfield  {author} {\bibinfo {author} {\bibfnamefont {P.~T.}\ \bibnamefont
  {Leung}}, \bibinfo {author} {\bibfnamefont {S.~Y.}\ \bibnamefont {Liu}}, \
  and\ \bibinfo {author} {\bibfnamefont {K.}~\bibnamefont {Young}},\
  }\href@noop {} {\bibfield  {journal} {\bibinfo  {journal} {Physical Review
  A}\ }\textbf {\bibinfo {volume} {49}},\ \bibinfo {pages} {3057} (\bibinfo
  {year} {1994}{\natexlab{a}})}\BibitemShut {NoStop}%
\bibitem [{\citenamefont {Ching}\ \emph {et~al.}(1996)\citenamefont {Ching},
  \citenamefont {Leung},\ and\ \citenamefont {Young}}]{Ching_1996}%
  \BibitemOpen
  \bibfield  {author} {\bibinfo {author} {\bibfnamefont {E.~S.-C.}\
  \bibnamefont {Ching}}, \bibinfo {author} {\bibfnamefont {P.-T.}\ \bibnamefont
  {Leung}}, \ and\ \bibinfo {author} {\bibfnamefont {K.}~\bibnamefont
  {Young}},\ }\href@noop {} {\emph {\bibinfo {title} {Optical processes in
  microcavities�the role of quasi-normal modes, R. K. Chang and A. J.
  Campillo Eds.}}}\ (\bibinfo  {publisher} {World Scientic},\ \bibinfo {year}
  {1996})\BibitemShut {NoStop}%
\bibitem [{\citenamefont {Leung}\ and\ \citenamefont
  {Pang}(1996)}]{Leung_JOSAB_13_805_1996}%
  \BibitemOpen
  \bibfield  {author} {\bibinfo {author} {\bibfnamefont {P.~T.}\ \bibnamefont
  {Leung}}\ and\ \bibinfo {author} {\bibfnamefont {K.~M.}\ \bibnamefont
  {Pang}},\ }\href@noop {} {\bibfield  {journal} {\bibinfo  {journal} {Journal
  of the Optical Society of America B}\ }\textbf {\bibinfo {volume} {13}},\
  \bibinfo {pages} {805} (\bibinfo {year} {1996})}\BibitemShut {NoStop}%
\bibitem [{\citenamefont {Lee}\ \emph {et~al.}(1999{\natexlab{a}})\citenamefont
  {Lee}, \citenamefont {Leung},\ and\ \citenamefont
  {Pang}}]{Lee_JOSAB_16_1409_1999}%
  \BibitemOpen
  \bibfield  {author} {\bibinfo {author} {\bibfnamefont {K.~M.}\ \bibnamefont
  {Lee}}, \bibinfo {author} {\bibfnamefont {P.~T.}\ \bibnamefont {Leung}}, \
  and\ \bibinfo {author} {\bibfnamefont {K.~M.}\ \bibnamefont {Pang}},\
  }\href@noop {} {\bibfield  {journal} {\bibinfo  {journal} {Journal of the
  Optical Society of America B}\ }\textbf {\bibinfo {volume} {16}},\ \bibinfo
  {pages} {1409} (\bibinfo {year} {1999}{\natexlab{a}})}\BibitemShut {NoStop}%
\bibitem [{\citenamefont {Lee}\ \emph {et~al.}(1999{\natexlab{b}})\citenamefont
  {Lee}, \citenamefont {Leung},\ and\ \citenamefont
  {Pang}}]{Lee_JOSAB_16_1418_1999}%
  \BibitemOpen
  \bibfield  {author} {\bibinfo {author} {\bibfnamefont {K.~M.}\ \bibnamefont
  {Lee}}, \bibinfo {author} {\bibfnamefont {P.~T.}\ \bibnamefont {Leung}}, \
  and\ \bibinfo {author} {\bibfnamefont {K.~M.}\ \bibnamefont {Pang}},\
  }\href@noop {} {\bibfield  {journal} {\bibinfo  {journal} {Journal of the
  Optical Society of America B}\ }\textbf {\bibinfo {volume} {16}},\ \bibinfo
  {pages} {1418} (\bibinfo {year} {1999}{\natexlab{b}})}\BibitemShut {NoStop}%
\bibitem [{\citenamefont {Mie}(1908)}]{Mie}%
  \BibitemOpen
  \bibfield  {author} {\bibinfo {author} {\bibfnamefont {G.}~\bibnamefont
  {Mie}},\ }\href@noop {} {\bibfield  {journal} {\bibinfo  {journal} {Annals of
  Physics}\ }\textbf {\bibinfo {volume} {25}},\ \bibinfo {pages} {377}
  (\bibinfo {year} {1908})}\BibitemShut {NoStop}%
\bibitem [{\citenamefont {T\"ureci}\ \emph {et~al.}(2006)\citenamefont
  {T\"ureci}, \citenamefont {Stone},\ and\ \citenamefont
  {Collier}}]{Tureci_PRA_74_043822_2006}%
  \BibitemOpen
  \bibfield  {author} {\bibinfo {author} {\bibfnamefont {H.~E.}\ \bibnamefont
  {T\"ureci}}, \bibinfo {author} {\bibfnamefont {A.~D.}\ \bibnamefont {Stone}},
  \ and\ \bibinfo {author} {\bibfnamefont {B.}~\bibnamefont {Collier}},\
  }\href@noop {} {\bibfield  {journal} {\bibinfo  {journal} {Physical Review
  A}\ }\textbf {\bibinfo {volume} {74}},\ \bibinfo {pages} {043822} (\bibinfo
  {year} {2006})}\BibitemShut {NoStop}%
\bibitem [{\citenamefont {Andreasen}\ \emph {et~al.}(2011)\citenamefont
  {Andreasen}, \citenamefont {Asatryan}, \citenamefont {Botten}, \citenamefont
  {Byrne}, \citenamefont {Cao}, \citenamefont {Ge}, \citenamefont
  {Labont\'{e}}, \citenamefont {Sebbah}, \citenamefont {Stone}, \citenamefont
  {T\"{u}reci},\ and\ \citenamefont
  {Vanneste}}]{Andreasen_AdvOptPhot_3_88_2011}%
  \BibitemOpen
  \bibfield  {author} {\bibinfo {author} {\bibfnamefont {J.}~\bibnamefont
  {Andreasen}}, \bibinfo {author} {\bibfnamefont {A.~A.}\ \bibnamefont
  {Asatryan}}, \bibinfo {author} {\bibfnamefont {L.~C.}\ \bibnamefont
  {Botten}}, \bibinfo {author} {\bibfnamefont {M.~A.}\ \bibnamefont {Byrne}},
  \bibinfo {author} {\bibfnamefont {H.}~\bibnamefont {Cao}}, \bibinfo {author}
  {\bibfnamefont {L.}~\bibnamefont {Ge}}, \bibinfo {author} {\bibfnamefont
  {L.}~\bibnamefont {Labont\'{e}}}, \bibinfo {author} {\bibfnamefont
  {P.}~\bibnamefont {Sebbah}}, \bibinfo {author} {\bibfnamefont {A.~D.}\
  \bibnamefont {Stone}}, \bibinfo {author} {\bibfnamefont {H.~E.}\ \bibnamefont
  {T\"{u}reci}}, \ and\ \bibinfo {author} {\bibfnamefont {C.}~\bibnamefont
  {Vanneste}},\ }\href@noop {} {\bibfield  {journal} {\bibinfo  {journal}
  {Advances in Optics and Photonics}\ }\textbf {\bibinfo {volume} {3}},\
  \bibinfo {pages} {88} (\bibinfo {year} {2011})}\BibitemShut {NoStop}%
\bibitem [{\citenamefont {Settimi}\ \emph {et~al.}(2003)\citenamefont
  {Settimi}, \citenamefont {Severini}, \citenamefont {Mattiucci}, \citenamefont
  {Sibilia}, \citenamefont {Centini}, \citenamefont {D'Aguanno}, \citenamefont
  {Bertolotti}, \citenamefont {Scalora}, \citenamefont {Bloemer},\ and\
  \citenamefont {Bowden}}]{Settimi_PRE_68_026614_2003}%
  \BibitemOpen
  \bibfield  {author} {\bibinfo {author} {\bibfnamefont {A.}~\bibnamefont
  {Settimi}}, \bibinfo {author} {\bibfnamefont {S.}~\bibnamefont {Severini}},
  \bibinfo {author} {\bibfnamefont {N.}~\bibnamefont {Mattiucci}}, \bibinfo
  {author} {\bibfnamefont {C.}~\bibnamefont {Sibilia}}, \bibinfo {author}
  {\bibfnamefont {M.}~\bibnamefont {Centini}}, \bibinfo {author} {\bibfnamefont
  {G.}~\bibnamefont {D'Aguanno}}, \bibinfo {author} {\bibfnamefont
  {M.}~\bibnamefont {Bertolotti}}, \bibinfo {author} {\bibfnamefont
  {M.}~\bibnamefont {Scalora}}, \bibinfo {author} {\bibfnamefont
  {M.}~\bibnamefont {Bloemer}}, \ and\ \bibinfo {author} {\bibfnamefont
  {C.~M.}\ \bibnamefont {Bowden}},\ }\href@noop {} {\bibfield  {journal}
  {\bibinfo  {journal} {Physical Review E}\ }\textbf {\bibinfo {volume} {68}},\
  \bibinfo {pages} {026614} (\bibinfo {year} {2003})}\BibitemShut {NoStop}%
\bibitem [{\citenamefont {Settimi}\ \emph {et~al.}(2009)\citenamefont
  {Settimi}, \citenamefont {Severini},\ and\ \citenamefont
  {Hoenders}}]{Settimi_JOSAB_26_876_2009}%
  \BibitemOpen
  \bibfield  {author} {\bibinfo {author} {\bibfnamefont {A.}~\bibnamefont
  {Settimi}}, \bibinfo {author} {\bibfnamefont {S.}~\bibnamefont {Severini}}, \
  and\ \bibinfo {author} {\bibfnamefont {B.~J.}\ \bibnamefont {Hoenders}},\
  }\href@noop {} {\bibfield  {journal} {\bibinfo  {journal} {Journal of the
  Optical Society of America B}\ }\textbf {\bibinfo {volume} {26}},\ \bibinfo
  {pages} {876} (\bibinfo {year} {2009})}\BibitemShut {NoStop}%
\bibitem [{\citenamefont {Maksimovi\'c}\ \emph {et~al.}(2008)\citenamefont
  {Maksimovi\'c}, \citenamefont {Hammer},\ and\ \citenamefont {van
  Groesen}}]{Maksimovic_OptEng_47_114601_2008}%
  \BibitemOpen
  \bibfield  {author} {\bibinfo {author} {\bibfnamefont {M.}~\bibnamefont
  {Maksimovi\'c}}, \bibinfo {author} {\bibfnamefont {M.}~\bibnamefont
  {Hammer}}, \ and\ \bibinfo {author} {\bibfnamefont {E.~W. C.~B.}\
  \bibnamefont {van Groesen}},\ }\href@noop {} {\bibfield  {journal} {\bibinfo
  {journal} {Optical Engineering}\ }\textbf {\bibinfo {volume} {47}},\ \bibinfo
  {pages} {114601} (\bibinfo {year} {2008})}\BibitemShut {NoStop}%
\bibitem [{\citenamefont {Ho}\ \emph {et~al.}(1998)\citenamefont {Ho},
  \citenamefont {Leung}, \citenamefont {Maassen van~den Brink},\ and\
  \citenamefont {Young}}]{Ho_PRE_58_2965_1998}%
  \BibitemOpen
  \bibfield  {author} {\bibinfo {author} {\bibfnamefont {K.~C.}\ \bibnamefont
  {Ho}}, \bibinfo {author} {\bibfnamefont {P.~T.}\ \bibnamefont {Leung}},
  \bibinfo {author} {\bibfnamefont {A.}~\bibnamefont {Maassen van~den Brink}},
  \ and\ \bibinfo {author} {\bibfnamefont {K.}~\bibnamefont {Young}},\
  }\href@noop {} {\bibfield  {journal} {\bibinfo  {journal} {Physical Review
  E}\ }\textbf {\bibinfo {volume} {58}},\ \bibinfo {pages} {2965} (\bibinfo
  {year} {1998})}\BibitemShut {NoStop}%
\bibitem [{\citenamefont {Dutra}\ and\ \citenamefont
  {Nienhuis}(2000)}]{Dutra_PRA_62_063805_2000}%
  \BibitemOpen
  \bibfield  {author} {\bibinfo {author} {\bibfnamefont {S.~M.}\ \bibnamefont
  {Dutra}}\ and\ \bibinfo {author} {\bibfnamefont {G.}~\bibnamefont
  {Nienhuis}},\ }\href@noop {} {\bibfield  {journal} {\bibinfo  {journal}
  {Physical Review A}\ }\textbf {\bibinfo {volume} {62}},\ \bibinfo {pages}
  {063805} (\bibinfo {year} {2000})}\BibitemShut {NoStop}%
\bibitem [{\citenamefont {Severini}\ \emph {et~al.}(2004)\citenamefont
  {Severini}, \citenamefont {Settimi}, \citenamefont {Sibilia}, \citenamefont
  {Bertolotti}, \citenamefont {Napoli},\ and\ \citenamefont
  {Messina}}]{Severini_PRE_70_056614_2004}%
  \BibitemOpen
  \bibfield  {author} {\bibinfo {author} {\bibfnamefont {S.}~\bibnamefont
  {Severini}}, \bibinfo {author} {\bibfnamefont {A.}~\bibnamefont {Settimi}},
  \bibinfo {author} {\bibfnamefont {C.}~\bibnamefont {Sibilia}}, \bibinfo
  {author} {\bibfnamefont {M.}~\bibnamefont {Bertolotti}}, \bibinfo {author}
  {\bibfnamefont {A.}~\bibnamefont {Napoli}}, \ and\ \bibinfo {author}
  {\bibfnamefont {A.}~\bibnamefont {Messina}},\ }\href@noop {} {\bibfield
  {journal} {\bibinfo  {journal} {Physical Review E}\ }\textbf {\bibinfo
  {volume} {70}},\ \bibinfo {pages} {056614} (\bibinfo {year}
  {2004})}\BibitemShut {NoStop}%
\bibitem [{\citenamefont {Dignam}\ and\ \citenamefont
  {Dezfouli}(2012)}]{Dignam_PRA_85_013809_2012}%
  \BibitemOpen
  \bibfield  {author} {\bibinfo {author} {\bibfnamefont {M.~M.}\ \bibnamefont
  {Dignam}}\ and\ \bibinfo {author} {\bibfnamefont {M.~K.}\ \bibnamefont
  {Dezfouli}},\ }\href@noop {} {\bibfield  {journal} {\bibinfo  {journal}
  {Physical Review A}\ }\textbf {\bibinfo {volume} {85}},\ \bibinfo {pages}
  {013809} (\bibinfo {year} {2012})}\BibitemShut {NoStop}%
\bibitem [{\citenamefont {Dignam}\ \emph {et~al.}(2006)\citenamefont {Dignam},
  \citenamefont {Fussell}, \citenamefont {Steel}, \citenamefont {de~Sterke},\
  and\ \citenamefont {McPhedran}}]{MarcQM1}%
  \BibitemOpen
  \bibfield  {author} {\bibinfo {author} {\bibfnamefont {M.~M.}\ \bibnamefont
  {Dignam}}, \bibinfo {author} {\bibfnamefont {D.~P.}\ \bibnamefont {Fussell}},
  \bibinfo {author} {\bibfnamefont {M.~J.}\ \bibnamefont {Steel}}, \bibinfo
  {author} {\bibfnamefont {C.~M.}\ \bibnamefont {de~Sterke}}, \ and\ \bibinfo
  {author} {\bibfnamefont {R.~C.}\ \bibnamefont {McPhedran}},\ }\href@noop {}
  {\bibfield  {journal} {\bibinfo  {journal} {Physical Review Letters}\
  }\textbf {\bibinfo {volume} {96}},\ \bibinfo {pages} {103902} (\bibinfo
  {year} {2006})}\BibitemShut {NoStop}%
\bibitem [{\citenamefont {Fussell}\ and\ \citenamefont
  {Dignam}(2008)}]{MarcQM2}%
  \BibitemOpen
  \bibfield  {author} {\bibinfo {author} {\bibfnamefont {D.~P.}\ \bibnamefont
  {Fussell}}\ and\ \bibinfo {author} {\bibfnamefont {M.~M.}\ \bibnamefont
  {Dignam}},\ }\href@noop {} {\bibfield  {journal} {\bibinfo  {journal}
  {Physical Review A}\ }\textbf {\bibinfo {volume} {77}},\ \bibinfo {pages}
  {053805} (\bibinfo {year} {2008})}\BibitemShut {NoStop}%
\bibitem [{\citenamefont {Siegert}(1939)}]{Siegert39}%
  \BibitemOpen
  \bibfield  {author} {\bibinfo {author} {\bibfnamefont {A.~J.~F.}\
  \bibnamefont {Siegert}},\ }\href@noop {} {\bibfield  {journal} {\bibinfo
  {journal} {Physical Review}\ }\textbf {\bibinfo {volume} {56}},\ \bibinfo
  {pages} {750} (\bibinfo {year} {1939})}\BibitemShut {NoStop}%
\bibitem [{\citenamefont {Santra}\ \emph {et~al.}(2005)\citenamefont {Santra},
  \citenamefont {Shainline},\ and\ \citenamefont {Greene}}]{Siegert05}%
  \BibitemOpen
  \bibfield  {author} {\bibinfo {author} {\bibfnamefont {R.}~\bibnamefont
  {Santra}}, \bibinfo {author} {\bibfnamefont {J.~M.}\ \bibnamefont
  {Shainline}}, \ and\ \bibinfo {author} {\bibfnamefont {C.~H.}\ \bibnamefont
  {Greene}},\ }\href@noop {} {\bibfield  {journal} {\bibinfo  {journal}
  {Physical Review A}\ }\textbf {\bibinfo {volume} {71}},\ \bibinfo {pages}
  {032703} (\bibinfo {year} {2005})}\BibitemShut {NoStop}%
\bibitem [{\citenamefont {Martin}(2006)}]{Martin_MultipleScattering}%
  \BibitemOpen
  \bibfield  {author} {\bibinfo {author} {\bibfnamefont {P.}~\bibnamefont
  {Martin}},\ }\href@noop {} {\emph {\bibinfo {title} {Multiple Scattering.
  Interaction of time-harmonic waves with N obstacles}}}\ (\bibinfo
  {publisher} {Cambridge University Press},\ \bibinfo {year}
  {2006})\BibitemShut {NoStop}%
\bibitem [{\citenamefont {Iserles}(2006)}]{Iserles_2003}%
  \BibitemOpen
  \bibfield  {author} {\bibinfo {author} {\bibfnamefont {A.}~\bibnamefont
  {Iserles}},\ }\href@noop {} {\emph {\bibinfo {title} {A First Course in the
  Numerical Analysis of Differential Equations}}}\ (\bibinfo  {publisher}
  {Cambridge University Press},\ \bibinfo {year} {2006})\BibitemShut {NoStop}%
\bibitem [{\citenamefont {Armaroli}\ \emph {et~al.}(2008)\citenamefont
  {Armaroli}, \citenamefont {Morand}, \citenamefont {Benech}, \citenamefont
  {Bellanca},\ and\ \citenamefont {Trillo}}]{Armaroli_JOSAA_25_667_2008}%
  \BibitemOpen
  \bibfield  {author} {\bibinfo {author} {\bibfnamefont {A.}~\bibnamefont
  {Armaroli}}, \bibinfo {author} {\bibfnamefont {A.}~\bibnamefont {Morand}},
  \bibinfo {author} {\bibfnamefont {P.}~\bibnamefont {Benech}}, \bibinfo
  {author} {\bibfnamefont {G.}~\bibnamefont {Bellanca}}, \ and\ \bibinfo
  {author} {\bibfnamefont {S.}~\bibnamefont {Trillo}},\ }\href@noop {}
  {\bibfield  {journal} {\bibinfo  {journal} {Journal of the Optical Society of
  America A}\ }\textbf {\bibinfo {volume} {25}},\ \bibinfo {pages} {667}
  (\bibinfo {year} {2008})}\BibitemShut {NoStop}%
\bibitem [{\citenamefont {Moharam}\ and\ \citenamefont
  {Gaylord}(1981)}]{Moharam_JOSA_71_811_1981}%
  \BibitemOpen
  \bibfield  {author} {\bibinfo {author} {\bibfnamefont {M.~G.}\ \bibnamefont
  {Moharam}}\ and\ \bibinfo {author} {\bibfnamefont {T.~K.}\ \bibnamefont
  {Gaylord}},\ }\href@noop {} {\bibfield  {journal} {\bibinfo  {journal}
  {Journal of the Optical Society of America}\ }\textbf {\bibinfo {volume}
  {71}},\ \bibinfo {pages} {811} (\bibinfo {year} {1981})}\BibitemShut
  {NoStop}%
\bibitem [{\citenamefont {Tavlove}(1995)}]{Tavlove_1995}%
  \BibitemOpen
  \bibfield  {author} {\bibinfo {author} {\bibfnamefont {A.}~\bibnamefont
  {Tavlove}},\ }\href@noop {} {\emph {\bibinfo {title} {Computational
  Electromagnetics: The Finite-difference time-domain method}}}\ (\bibinfo
  {publisher} {Artech House},\ \bibinfo {year} {1995})\BibitemShut {NoStop}%
\bibitem [{\citenamefont {Maes}\ \emph {et~al.}(2013)\citenamefont {Maes},
  \citenamefont {Petr\'{a}\v{c}ek}, \citenamefont {Burger}, \citenamefont
  {Kwiecien}, \citenamefont {Luksch},\ and\ \citenamefont
  {Richter}}]{Maes_OE_21_6794_2013}%
  \BibitemOpen
  \bibfield  {author} {\bibinfo {author} {\bibfnamefont {B.}~\bibnamefont
  {Maes}}, \bibinfo {author} {\bibfnamefont {J.}~\bibnamefont
  {Petr\'{a}\v{c}ek}}, \bibinfo {author} {\bibfnamefont {S.}~\bibnamefont
  {Burger}}, \bibinfo {author} {\bibfnamefont {P.}~\bibnamefont {Kwiecien}},
  \bibinfo {author} {\bibfnamefont {J.}~\bibnamefont {Luksch}}, \ and\ \bibinfo
  {author} {\bibfnamefont {I.}~\bibnamefont {Richter}},\ }\href@noop {}
  {\bibfield  {journal} {\bibinfo  {journal} {Optics Express}\ }\textbf
  {\bibinfo {volume} {21}},\ \bibinfo {pages} {6794} (\bibinfo {year}
  {2013})}\BibitemShut {NoStop}%
\bibitem [{\citenamefont {Lalanne}\ \emph {et~al.}(2008)\citenamefont
  {Lalanne}, \citenamefont {Sauvan},\ and\ \citenamefont
  {Hugonin}}]{Lalanne_LaserPhotRev_2_514_2008}%
  \BibitemOpen
  \bibfield  {author} {\bibinfo {author} {\bibfnamefont {P.}~\bibnamefont
  {Lalanne}}, \bibinfo {author} {\bibfnamefont {C.}~\bibnamefont {Sauvan}}, \
  and\ \bibinfo {author} {\bibfnamefont {J.}~\bibnamefont {Hugonin}},\
  }\href@noop {} {\bibfield  {journal} {\bibinfo  {journal} {Laser \& Photonics
  Reviews}\ }\textbf {\bibinfo {volume} {2}},\ \bibinfo {pages} {514} (\bibinfo
  {year} {2008})}\BibitemShut {NoStop}%
\bibitem [{\citenamefont {Bryant}\ \emph {et~al.}(2008)\citenamefont {Bryant},
  \citenamefont {de~Abajo},\ and\ \citenamefont
  {Aizpurua}}]{Bryant_NL_8_631_2008}%
  \BibitemOpen
  \bibfield  {author} {\bibinfo {author} {\bibfnamefont {G.~W.}\ \bibnamefont
  {Bryant}}, \bibinfo {author} {\bibfnamefont {F.~J.~G.}\ \bibnamefont
  {de~Abajo}}, \ and\ \bibinfo {author} {\bibfnamefont {J.}~\bibnamefont
  {Aizpurua}},\ }\href@noop {} {\bibfield  {journal} {\bibinfo  {journal} {Nano
  Letters}\ }\textbf {\bibinfo {volume} {8}},\ \bibinfo {pages} {631} (\bibinfo
  {year} {2008})}\BibitemShut {NoStop}%
\bibitem [{\citenamefont {Taminiau}\ \emph {et~al.}(2011)\citenamefont
  {Taminiau}, \citenamefont {Stefani},\ and\ \citenamefont {van
  Hulst}}]{Taminiau_NL_11_1020_2011}%
  \BibitemOpen
  \bibfield  {author} {\bibinfo {author} {\bibfnamefont {T.~H.}\ \bibnamefont
  {Taminiau}}, \bibinfo {author} {\bibfnamefont {F.~D.}\ \bibnamefont
  {Stefani}}, \ and\ \bibinfo {author} {\bibfnamefont {N.~F.}\ \bibnamefont
  {van Hulst}},\ }\href@noop {} {\bibfield  {journal} {\bibinfo  {journal}
  {Nano Letters}\ }\textbf {\bibinfo {volume} {11}},\ \bibinfo {pages} {1020}
  (\bibinfo {year} {2011})}\BibitemShut {NoStop}%
\bibitem [{\citenamefont {Leung}\ \emph
  {et~al.}(1994{\natexlab{b}})\citenamefont {Leung}, \citenamefont {Liu},\ and\
  \citenamefont {Young}}]{Leung_PRA_49_3982}%
  \BibitemOpen
  \bibfield  {author} {\bibinfo {author} {\bibfnamefont {P.~T.}\ \bibnamefont
  {Leung}}, \bibinfo {author} {\bibfnamefont {S.~Y.}\ \bibnamefont {Liu}}, \
  and\ \bibinfo {author} {\bibfnamefont {K.}~\bibnamefont {Young}},\
  }\href@noop {} {\bibfield  {journal} {\bibinfo  {journal} {Physical Review
  A}\ }\textbf {\bibinfo {volume} {49}},\ \bibinfo {pages} {3982} (\bibinfo
  {year} {1994}{\natexlab{b}})}\BibitemShut {NoStop}%
\bibitem [{\citenamefont {Sauvan}\ \emph {et~al.}(2013)\citenamefont {Sauvan},
  \citenamefont {Hugonin}, \citenamefont {Maksymov},\ and\ \citenamefont
  {Lalanne}}]{Sauvan_PRL_110_237401_2013}%
  \BibitemOpen
  \bibfield  {author} {\bibinfo {author} {\bibfnamefont {C.}~\bibnamefont
  {Sauvan}}, \bibinfo {author} {\bibfnamefont {J.~P.}\ \bibnamefont {Hugonin}},
  \bibinfo {author} {\bibfnamefont {I.~S.}\ \bibnamefont {Maksymov}}, \ and\
  \bibinfo {author} {\bibfnamefont {P.}~\bibnamefont {Lalanne}},\ }\href@noop
  {} {\bibfield  {journal} {\bibinfo  {journal} {Physical Review Letters}\
  }\textbf {\bibinfo {volume} {110}},\ \bibinfo {pages} {237401} (\bibinfo
  {year} {2013})}\BibitemShut {NoStop}%
\bibitem [{\citenamefont {Snyder}\ and\ \citenamefont
  {Love}(1983)}]{SnyderLoveBook}%
  \BibitemOpen
  \bibfield  {author} {\bibinfo {author} {\bibfnamefont {A.~W.}\ \bibnamefont
  {Snyder}}\ and\ \bibinfo {author} {\bibfnamefont {J.}~\bibnamefont {Love}},\
  }\href@noop {} {\emph {\bibinfo {title} {Optical Waveguide Theory}}}\
  (\bibinfo  {publisher} {Kluwar Academic Publishers},\ \bibinfo {year}
  {1983})\BibitemShut {NoStop}%
\bibitem [{\citenamefont {Breukelaar}\ \emph {et~al.}(2006)\citenamefont
  {Breukelaar}, \citenamefont {Charbonneau},\ and\ \citenamefont
  {Berini}}]{Berini2006}%
  \BibitemOpen
  \bibfield  {author} {\bibinfo {author} {\bibfnamefont {I.}~\bibnamefont
  {Breukelaar}}, \bibinfo {author} {\bibfnamefont {R.}~\bibnamefont
  {Charbonneau}}, \ and\ \bibinfo {author} {\bibfnamefont {P.}~\bibnamefont
  {Berini}},\ }\href@noop {} {\bibfield  {journal} {\bibinfo  {journal}
  {Journal of Applied Physics}\ }\textbf {\bibinfo {volume} {100}},\ \bibinfo
  {pages} {043104} (\bibinfo {year} {2006})}\BibitemShut {NoStop}%
\bibitem [{\citenamefont {Bai}\ \emph {et~al.}(2013)\citenamefont {Bai},
  \citenamefont {Perrin}, \citenamefont {Sauvan}, \citenamefont {Hugonin},\
  and\ \citenamefont {Lalanne}}]{Bai_OE_21_27371_2013}%
  \BibitemOpen
  \bibfield  {author} {\bibinfo {author} {\bibfnamefont {Q.}~\bibnamefont
  {Bai}}, \bibinfo {author} {\bibfnamefont {M.}~\bibnamefont {Perrin}},
  \bibinfo {author} {\bibfnamefont {C.}~\bibnamefont {Sauvan}}, \bibinfo
  {author} {\bibfnamefont {J.-P.}\ \bibnamefont {Hugonin}}, \ and\ \bibinfo
  {author} {\bibfnamefont {P.}~\bibnamefont {Lalanne}},\ }\href@noop {}
  {\bibfield  {journal} {\bibinfo  {journal} {Optics Express}\ }\textbf
  {\bibinfo {volume} {21}},\ \bibinfo {pages} {27371} (\bibinfo {year}
  {2013})}\BibitemShut {NoStop}%
\bibitem [{\citenamefont {Tai}(1994)}]{Tai_1994}%
  \BibitemOpen
  \bibfield  {author} {\bibinfo {author} {\bibfnamefont {C.-T.}\ \bibnamefont
  {Tai}},\ }\href@noop {} {\emph {\bibinfo {title} {Dyadic Green Functions in
  Electromagnetic Theory, 2nd ed.}}}\ (\bibinfo  {publisher} {IEEE Press},\
  \bibinfo {year} {1994})\BibitemShut {NoStop}%
\bibitem [{\citenamefont {Martin}\ and\ \citenamefont
  {Piller}(1998)}]{Martin_PRE_58_3909_1998}%
  \BibitemOpen
  \bibfield  {author} {\bibinfo {author} {\bibfnamefont {O.~J.~F.}\
  \bibnamefont {Martin}}\ and\ \bibinfo {author} {\bibfnamefont {N.~B.}\
  \bibnamefont {Piller}},\ }\href@noop {} {\bibfield  {journal} {\bibinfo
  {journal} {Physical Review E}\ }\textbf {\bibinfo {volume} {58}},\ \bibinfo
  {pages} {3909} (\bibinfo {year} {1998})}\BibitemShut {NoStop}%
\bibitem [{\citenamefont {Novotny}\ and\ \citenamefont
  {Hecht}(2006)}]{NovotnyAndHecht_2006}%
  \BibitemOpen
  \bibfield  {author} {\bibinfo {author} {\bibfnamefont {L.}~\bibnamefont
  {Novotny}}\ and\ \bibinfo {author} {\bibfnamefont {B.}~\bibnamefont
  {Hecht}},\ }\href@noop {} {\emph {\bibinfo {title} {Principles of Nano
  Optics}}}\ (\bibinfo  {publisher} {Cambridge University Press},\ \bibinfo
  {year} {2006})\BibitemShut {NoStop}%
\bibitem [{\citenamefont {Sprik}\ \emph {et~al.}(1996)\citenamefont {Sprik},
  \citenamefont {van Tiggelen},\ and\ \citenamefont
  {Lagendijk}}]{Sprik_EuroPhys_35_265_1996}%
  \BibitemOpen
  \bibfield  {author} {\bibinfo {author} {\bibfnamefont {R.}~\bibnamefont
  {Sprik}}, \bibinfo {author} {\bibfnamefont {B.~A.}\ \bibnamefont {van
  Tiggelen}}, \ and\ \bibinfo {author} {\bibfnamefont {A.}~\bibnamefont
  {Lagendijk}},\ }\href@noop {} {\bibfield  {journal} {\bibinfo  {journal}
  {Europhysics Letters}\ }\textbf {\bibinfo {volume} {35}},\ \bibinfo {pages}
  {265} (\bibinfo {year} {1996})}\BibitemShut {NoStop}%
\bibitem [{\citenamefont {Yao}\ \emph {et~al.}(2010)\citenamefont {Yao},
  \citenamefont {Rao},\ and\ \citenamefont {Hughes}}]{YaoLaserReviews}%
  \BibitemOpen
  \bibfield  {author} {\bibinfo {author} {\bibfnamefont {P.}~\bibnamefont
  {Yao}}, \bibinfo {author} {\bibfnamefont {V.~S. C.~M.}\ \bibnamefont {Rao}},
  \ and\ \bibinfo {author} {\bibfnamefont {S.}~\bibnamefont {Hughes}},\
  }\href@noop {} {\bibfield  {journal} {\bibinfo  {journal} {Laser and
  Photonics Reviews}\ }\textbf {\bibinfo {volume} {4}},\ \bibinfo {pages} {499}
  (\bibinfo {year} {2010})}\BibitemShut {NoStop}%
\bibitem [{\citenamefont {Kristensen}\ \emph {et~al.}(2013)\citenamefont
  {Kristensen}, \citenamefont {Mortensen}, \citenamefont {Lodahl},\ and\
  \citenamefont {Stobbe}}]{Kristensen_PRB_88_205308_2013}%
  \BibitemOpen
  \bibfield  {author} {\bibinfo {author} {\bibfnamefont {P.~T.}\ \bibnamefont
  {Kristensen}}, \bibinfo {author} {\bibfnamefont {J.~E.}\ \bibnamefont
  {Mortensen}}, \bibinfo {author} {\bibfnamefont {P.}~\bibnamefont {Lodahl}}, \
  and\ \bibinfo {author} {\bibfnamefont {S.}~\bibnamefont {Stobbe}},\
  }\href@noop {} {\bibfield  {journal} {\bibinfo  {journal} {Physical Review
  B}\ }\textbf {\bibinfo {volume} {88}},\ \bibinfo {pages} {205308} (\bibinfo
  {year} {2013})}\BibitemShut {NoStop}%
\bibitem [{\citenamefont {Purcell}(1946)}]{Purcell_PR_69_681_1946}%
  \BibitemOpen
  \bibfield  {author} {\bibinfo {author} {\bibfnamefont {E.~M.}\ \bibnamefont
  {Purcell}},\ }\href@noop {} {\bibfield  {journal} {\bibinfo  {journal}
  {Physical Review}\ }\textbf {\bibinfo {volume} {69}},\ \bibinfo {pages} {681}
  (\bibinfo {year} {1946})}\BibitemShut {NoStop}%
\bibitem [{\citenamefont {Ge}\ \emph {et~al.}(2013{\natexlab{b}})\citenamefont
  {Ge}, \citenamefont {Kristensen}, \citenamefont {Young},\ and\ \citenamefont
  {Hughes}}]{oursMNP}%
  \BibitemOpen
  \bibfield  {author} {\bibinfo {author} {\bibfnamefont {R.-C.}\ \bibnamefont
  {Ge}}, \bibinfo {author} {\bibfnamefont {P.~T.}\ \bibnamefont {Kristensen}},
  \bibinfo {author} {\bibfnamefont {J.~F.}\ \bibnamefont {Young}}, \ and\
  \bibinfo {author} {\bibfnamefont {S.}~\bibnamefont {Hughes}},\ }\href@noop {}
  {\bibfield  {journal} {\bibinfo  {journal} {arXiv:1312.2939 [quant-ph]}\ }
  (\bibinfo {year} {2013}{\natexlab{b}})}\BibitemShut {NoStop}%
\bibitem [{\citenamefont {G\'{e}rard}\ and\ \citenamefont
  {Gayral}(1999)}]{Gerard_JLT_17_2089_1999}%
  \BibitemOpen
  \bibfield  {author} {\bibinfo {author} {\bibfnamefont {J.-M.}\ \bibnamefont
  {G\'{e}rard}}\ and\ \bibinfo {author} {\bibfnamefont {B.}~\bibnamefont
  {Gayral}},\ }\href@noop {} {\bibfield  {journal} {\bibinfo  {journal} {IEEE
  Journal of Lightwave Technology}\ }\textbf {\bibinfo {volume} {17}},\
  \bibinfo {pages} {2089} (\bibinfo {year} {1999})}\BibitemShut {NoStop}%
\bibitem [{\citenamefont {Maier}(2006)}]{MaierOE}%
  \BibitemOpen
  \bibfield  {author} {\bibinfo {author} {\bibfnamefont {S.~A.}\ \bibnamefont
  {Maier}},\ }\href@noop {} {\bibfield  {journal} {\bibinfo  {journal} {Optics
  Express}\ }\textbf {\bibinfo {volume} {14}},\ \bibinfo {pages} {1957}
  (\bibinfo {year} {2006})}\BibitemShut {NoStop}%
\bibitem [{\citenamefont {Ruppin}(2012)}]{Ruppin02}%
  \BibitemOpen
  \bibfield  {author} {\bibinfo {author} {\bibfnamefont {R.}~\bibnamefont
  {Ruppin}},\ }\href@noop {} {\bibfield  {journal} {\bibinfo  {journal}
  {Physics Letters A}\ }\textbf {\bibinfo {volume} {299}},\ \bibinfo {pages}
  {309} (\bibinfo {year} {2012})}\BibitemShut {NoStop}%
\bibitem [{\citenamefont {Koenderink}(2010)}]{Koenderink_OL_35_4208_2010}%
  \BibitemOpen
  \bibfield  {author} {\bibinfo {author} {\bibfnamefont {A.~F.}\ \bibnamefont
  {Koenderink}},\ }\href@noop {} {\bibfield  {journal} {\bibinfo  {journal}
  {Optics Letters}\ }\textbf {\bibinfo {volume} {35}},\ \bibinfo {pages} {4208}
  (\bibinfo {year} {2010})}\BibitemShut {NoStop}%
\bibitem [{Lum()}]{Lumerical}%
  \BibitemOpen
  \href@noop {} {}\bibinfo {note} {We used ``FDTD Solutions'' from Lumerical
  Solutions: www.lumerical.com}\BibitemShut {NoStop}%
\bibitem [{\citenamefont {Liscidini}\ \emph {et~al.}(2012)\citenamefont
  {Liscidini}, \citenamefont {Helt},\ and\ \citenamefont
  {Sipe}}]{Liscidini_PRA_85_013833_2012}%
  \BibitemOpen
  \bibfield  {author} {\bibinfo {author} {\bibfnamefont {M.}~\bibnamefont
  {Liscidini}}, \bibinfo {author} {\bibfnamefont {L.~G.}\ \bibnamefont {Helt}},
  \ and\ \bibinfo {author} {\bibfnamefont {J.~E.}\ \bibnamefont {Sipe}},\
  }\href@noop {} {\bibfield  {journal} {\bibinfo  {journal} {Physical Review
  A}\ }\textbf {\bibinfo {volume} {85}},\ \bibinfo {pages} {013833} (\bibinfo
  {year} {2012})}\BibitemShut {NoStop}%
\bibitem [{\citenamefont {Manga~Rao}\ and\ \citenamefont
  {Hughes}(2007)}]{MangaRao_PRL_99_193901_2007}%
  \BibitemOpen
  \bibfield  {author} {\bibinfo {author} {\bibfnamefont {V.~S.~C.}\
  \bibnamefont {Manga~Rao}}\ and\ \bibinfo {author} {\bibfnamefont
  {S.}~\bibnamefont {Hughes}},\ }\href@noop {} {\bibfield  {journal} {\bibinfo
  {journal} {Physical Review Letters}\ }\textbf {\bibinfo {volume} {99}},\
  \bibinfo {pages} {193901} (\bibinfo {year} {2007})}\BibitemShut {NoStop}%
\bibitem [{\citenamefont {Cowan}\ and\ \citenamefont
  {Young}(2003)}]{Cowan_PRE_68_046606_2003}%
  \BibitemOpen
  \bibfield  {author} {\bibinfo {author} {\bibfnamefont {A.~R.}\ \bibnamefont
  {Cowan}}\ and\ \bibinfo {author} {\bibfnamefont {J.~F.}\ \bibnamefont
  {Young}},\ }\href@noop {} {\bibfield  {journal} {\bibinfo  {journal}
  {Physical Review E}\ }\textbf {\bibinfo {volume} {68}},\ \bibinfo {pages}
  {046606} (\bibinfo {year} {2003})}\BibitemShut {NoStop}%
\bibitem [{\citenamefont {Yao}\ and\ \citenamefont
  {Hughes}(2009{\natexlab{a}})}]{Yao_PRB_80_165128_2009}%
  \BibitemOpen
  \bibfield  {author} {\bibinfo {author} {\bibfnamefont {P.}~\bibnamefont
  {Yao}}\ and\ \bibinfo {author} {\bibfnamefont {S.}~\bibnamefont {Hughes}},\
  }\href@noop {} {\bibfield  {journal} {\bibinfo  {journal} {Physical Review
  B}\ }\textbf {\bibinfo {volume} {80}},\ \bibinfo {pages} {165128} (\bibinfo
  {year} {2009}{\natexlab{a}})}\BibitemShut {NoStop}%
\bibitem [{\citenamefont {Hughes}(2005)}]{Hughes_PRL_94_227402_2005}%
  \BibitemOpen
  \bibfield  {author} {\bibinfo {author} {\bibfnamefont {S.}~\bibnamefont
  {Hughes}},\ }\href@noop {} {\bibfield  {journal} {\bibinfo  {journal}
  {Physical Review Letters}\ }\textbf {\bibinfo {volume} {94}},\ \bibinfo
  {pages} {227402} (\bibinfo {year} {2005})}\BibitemShut {NoStop}%
\bibitem [{\citenamefont {Yao}\ and\ \citenamefont
  {Hughes}(2009{\natexlab{b}})}]{Yao_OE_17_11505_2009}%
  \BibitemOpen
  \bibfield  {author} {\bibinfo {author} {\bibfnamefont {P.}~\bibnamefont
  {Yao}}\ and\ \bibinfo {author} {\bibfnamefont {S.}~\bibnamefont {Hughes}},\
  }\href@noop {} {\bibfield  {journal} {\bibinfo  {journal} {Optics Express}\
  }\textbf {\bibinfo {volume} {17}},\ \bibinfo {pages} {11505} (\bibinfo {year}
  {2009}{\natexlab{b}})}\BibitemShut {NoStop}%
\bibitem [{\citenamefont {Kristensen}\ \emph {et~al.}(2011)\citenamefont
  {Kristensen}, \citenamefont {M{\o}rk}, \citenamefont {Lodahl},\ and\
  \citenamefont {Hughes}}]{Kristensen_PRB_83_075305_2011}%
  \BibitemOpen
  \bibfield  {author} {\bibinfo {author} {\bibfnamefont {P.~T.}\ \bibnamefont
  {Kristensen}}, \bibinfo {author} {\bibfnamefont {J.}~\bibnamefont {M{\o}rk}},
  \bibinfo {author} {\bibfnamefont {P.}~\bibnamefont {Lodahl}}, \ and\ \bibinfo
  {author} {\bibfnamefont {S.}~\bibnamefont {Hughes}},\ }\href@noop {}
  {\bibfield  {journal} {\bibinfo  {journal} {Physical Review B}\ }\textbf
  {\bibinfo {volume} {83}},\ \bibinfo {pages} {075305} (\bibinfo {year}
  {2011})}\BibitemShut {NoStop}%
\bibitem [{\citenamefont {Pierce}(1954)}]{Pierce_JAP_25_179_1954}%
  \BibitemOpen
  \bibfield  {author} {\bibinfo {author} {\bibfnamefont {J.~R.}\ \bibnamefont
  {Pierce}},\ }\href@noop {} {\bibfield  {journal} {\bibinfo  {journal}
  {Journal of Applied Physics}\ }\textbf {\bibinfo {volume} {25}},\ \bibinfo
  {pages} {179} (\bibinfo {year} {1954})}\BibitemShut {NoStop}%
\bibitem [{\citenamefont {Haus}\ and\ \citenamefont
  {Huang}(1991)}]{Haus_Proc_IEEE_79_1505_1991}%
  \BibitemOpen
  \bibfield  {author} {\bibinfo {author} {\bibfnamefont {H.~A.}\ \bibnamefont
  {Haus}}\ and\ \bibinfo {author} {\bibfnamefont {W.}~\bibnamefont {Huang}},\
  }\href@noop {} {\bibfield  {journal} {\bibinfo  {journal} {Proceedings of the
  IEEE}\ }\textbf {\bibinfo {volume} {79}},\ \bibinfo {pages} {1505} (\bibinfo
  {year} {1991})}\BibitemShut {NoStop}%
\bibitem [{\citenamefont {Haus}(1984)}]{Haus_1984}%
  \BibitemOpen
  \bibfield  {author} {\bibinfo {author} {\bibfnamefont {H.~A.}\ \bibnamefont
  {Haus}},\ }\href@noop {} {\emph {\bibinfo {title} {Waves and fields in
  optoelectronics}}}\ (\bibinfo  {publisher} {Prentice Hall},\ \bibinfo {year}
  {1984})\BibitemShut {NoStop}%
\bibitem [{\citenamefont {Joannopoulos}\ \emph {et~al.}(2008)\citenamefont
  {Joannopoulos}, \citenamefont {Johnson}, \citenamefont {Winn},\ and\
  \citenamefont {Meade}}]{Joannopoulos2008}%
  \BibitemOpen
  \bibfield  {author} {\bibinfo {author} {\bibfnamefont {J.~D.}\ \bibnamefont
  {Joannopoulos}}, \bibinfo {author} {\bibfnamefont {S.~G.}\ \bibnamefont
  {Johnson}}, \bibinfo {author} {\bibfnamefont {J.~N.}\ \bibnamefont {Winn}}, \
  and\ \bibinfo {author} {\bibfnamefont {R.~D.}\ \bibnamefont {Meade}},\
  }\href@noop {} {\emph {\bibinfo {title} {Photonic Crystals - Molding the Flow
  of Light, second edition}}}\ (\bibinfo  {publisher} {Princeton University
  Press},\ \bibinfo {year} {2008})\BibitemShut {NoStop}%
\bibitem [{\citenamefont {Heuck}\ \emph {et~al.}(2012)\citenamefont {Heuck},
  \citenamefont {Kristensen},\ and\ \citenamefont
  {M{\o}rk}}]{Heuck_CLEO_2012_JW4A-6}%
  \BibitemOpen
  \bibfield  {author} {\bibinfo {author} {\bibfnamefont {M.}~\bibnamefont
  {Heuck}}, \bibinfo {author} {\bibfnamefont {P.~T.}\ \bibnamefont
  {Kristensen}}, \ and\ \bibinfo {author} {\bibfnamefont {J.}~\bibnamefont
  {M{\o}rk}},\ }in\ \href@noop {} {\emph {\bibinfo {booktitle} {Conference on
  Lasers and Electro-Optics}}}\ (\bibinfo {year} {2012})\ pp.\ \bibinfo {pages}
  {JW4A--6}\BibitemShut {NoStop}%
\bibitem [{\citenamefont {Tr\o{}st~Kristensen}\ \emph
  {et~al.}(2013)\citenamefont {Tr\o{}st~Kristensen}, \citenamefont {Heuck},\
  and\ \citenamefont {M\o{}rk}}]{Kristensen_APL_102_041107_2013}%
  \BibitemOpen
  \bibfield  {author} {\bibinfo {author} {\bibfnamefont {P.}~\bibnamefont
  {Tr\o{}st~Kristensen}}, \bibinfo {author} {\bibfnamefont {M.}~\bibnamefont
  {Heuck}}, \ and\ \bibinfo {author} {\bibfnamefont {J.}~\bibnamefont
  {M\o{}rk}},\ }\href@noop {} {\bibfield  {journal} {\bibinfo  {journal}
  {Applied Physics Letters}\ }\textbf {\bibinfo {volume} {102}},\ \bibinfo
  {pages} {041107} (\bibinfo {year} {2013})}\BibitemShut {NoStop}%
\bibitem [{\citenamefont {Cirac{\`i}}\ \emph {et~al.}(2012)\citenamefont
  {Cirac{\`i}}, \citenamefont {Hill}, \citenamefont {Mock}, \citenamefont
  {Urzhumov}, \citenamefont {Fern{\'a}ndez-Dom{\'i}nguez}, \citenamefont
  {Maier}, \citenamefont {Pendry}, \citenamefont {Chilkoti},\ and\
  \citenamefont {Smith}}]{Ciraci_Science_337_1072_2012}%
  \BibitemOpen
  \bibfield  {author} {\bibinfo {author} {\bibfnamefont {C.}~\bibnamefont
  {Cirac{\`i}}}, \bibinfo {author} {\bibfnamefont {R.~T.}\ \bibnamefont
  {Hill}}, \bibinfo {author} {\bibfnamefont {J.~J.}\ \bibnamefont {Mock}},
  \bibinfo {author} {\bibfnamefont {Y.}~\bibnamefont {Urzhumov}}, \bibinfo
  {author} {\bibfnamefont {A.~I.}\ \bibnamefont {Fern{\'a}ndez-Dom{\'i}nguez}},
  \bibinfo {author} {\bibfnamefont {S.~A.}\ \bibnamefont {Maier}}, \bibinfo
  {author} {\bibfnamefont {J.~B.}\ \bibnamefont {Pendry}}, \bibinfo {author}
  {\bibfnamefont {A.}~\bibnamefont {Chilkoti}}, \ and\ \bibinfo {author}
  {\bibfnamefont {D.~R.}\ \bibnamefont {Smith}},\ }\href@noop {} {\bibfield
  {journal} {\bibinfo  {journal} {Science}\ }\textbf {\bibinfo {volume}
  {337}},\ \bibinfo {pages} {1072} (\bibinfo {year} {2012})}\BibitemShut
  {NoStop}%
\bibitem [{\citenamefont {Fern{\'a}ndez-Dom{\'i}nguez}\ \emph
  {et~al.}(2012)\citenamefont {Fern{\'a}ndez-Dom{\'i}nguez}, \citenamefont
  {Wiener}, \citenamefont {Garc{\'i}a-Vidal}, \citenamefont {Maier},\ and\
  \citenamefont {Pendry}}]{Fernandez-Dom_PRL_108_106802_2012}%
  \BibitemOpen
  \bibfield  {author} {\bibinfo {author} {\bibfnamefont {A.~I.}\ \bibnamefont
  {Fern{\'a}ndez-Dom{\'i}nguez}}, \bibinfo {author} {\bibfnamefont
  {A.}~\bibnamefont {Wiener}}, \bibinfo {author} {\bibfnamefont {F.~J.}\
  \bibnamefont {Garc{\'i}a-Vidal}}, \bibinfo {author} {\bibfnamefont {S.~A.}\
  \bibnamefont {Maier}}, \ and\ \bibinfo {author} {\bibfnamefont {J.~B.}\
  \bibnamefont {Pendry}},\ }\href@noop {} {\bibfield  {journal} {\bibinfo
  {journal} {Physical Review Letters}\ }\textbf {\bibinfo {volume} {108}},\
  \bibinfo {pages} {106802} (\bibinfo {year} {2012})}\BibitemShut {NoStop}%
\bibitem [{\citenamefont {Raza}\ \emph {et~al.}(2013)\citenamefont {Raza},
  \citenamefont {Yan}, \citenamefont {Stenger}, \citenamefont {Wubs},\ and\
  \citenamefont {Mortensen}}]{Raza_OE_21_27344_2013}%
  \BibitemOpen
  \bibfield  {author} {\bibinfo {author} {\bibfnamefont {S.}~\bibnamefont
  {Raza}}, \bibinfo {author} {\bibfnamefont {W.}~\bibnamefont {Yan}}, \bibinfo
  {author} {\bibfnamefont {N.}~\bibnamefont {Stenger}}, \bibinfo {author}
  {\bibfnamefont {M.}~\bibnamefont {Wubs}}, \ and\ \bibinfo {author}
  {\bibfnamefont {N.~A.}\ \bibnamefont {Mortensen}},\ }\href@noop {} {\bibfield
   {journal} {\bibinfo  {journal} {Optics Express}\ }\textbf {\bibinfo {volume}
  {21}},\ \bibinfo {pages} {27344} (\bibinfo {year} {2013})}\BibitemShut
  {NoStop}%
\bibitem [{\citenamefont {Toscano}\ \emph {et~al.}(2013)\citenamefont
  {Toscano}, \citenamefont {Raza}, \citenamefont {Yan}, \citenamefont
  {Jeppesen}, \citenamefont {Xiao}, \citenamefont {Wubs}, \citenamefont
  {Jauho}, \citenamefont {Bozhevolnyi},\ and\ \citenamefont
  {Mortensen}}]{Toscano_NanoPhot_2_161_2013}%
  \BibitemOpen
  \bibfield  {author} {\bibinfo {author} {\bibfnamefont {G.}~\bibnamefont
  {Toscano}}, \bibinfo {author} {\bibfnamefont {S.}~\bibnamefont {Raza}},
  \bibinfo {author} {\bibfnamefont {W.}~\bibnamefont {Yan}}, \bibinfo {author}
  {\bibfnamefont {C.}~\bibnamefont {Jeppesen}}, \bibinfo {author}
  {\bibfnamefont {S.}~\bibnamefont {Xiao}}, \bibinfo {author} {\bibfnamefont
  {M.}~\bibnamefont {Wubs}}, \bibinfo {author} {\bibfnamefont {A.-P.}\
  \bibnamefont {Jauho}}, \bibinfo {author} {\bibfnamefont {S.~I.}\ \bibnamefont
  {Bozhevolnyi}}, \ and\ \bibinfo {author} {\bibfnamefont {N.~A.}\ \bibnamefont
  {Mortensen}},\ }\href@noop {} {\bibfield  {journal} {\bibinfo  {journal}
  {Nanophotonics}\ }\textbf {\bibinfo {volume} {2}},\ \bibinfo {pages} {161}
  (\bibinfo {year} {2013})}\BibitemShut {NoStop}%
\end{thebibliography}%

\end{document}